\documentclass[aps,twocolumn,showpacs,superscriptaddress,showkeys,noeprint,groupedaddress,nolongbibliography]{revtex4-2}  % for review and submission
\usepackage{graphicx}  % needed for figures
\usepackage{dcolumn}   % needed for some tables
\usepackage{bm}        % for math
\usepackage{amsmath,amssymb,framed,amsthm,xcolor}   % for math
\usepackage{array,multirow,graphicx}

\def\apjs{Astrophys.~J.~Supp.~Ser.}
\def\apj{Astrophys.~J.}
\def\apjl{Astrophys.~J.~Lett.}
\def\aap{Astron.~Astrophys.}
\def\mnras{Mon.~Not.~R.~Astron.~Soc.}

\def\pre{Phys.~Rev.~E}
\def\prl{Phys.~Rev.~Lett.}

\def\pop{Phys.~Plasmas}
\def\pof{Phys.~Fluids}

\def\jpp{J.~Plasma Phys.}
\def\jfm{J.~Fluid Mech.}
\def\araa{Ann.~Rev.~Astron.~Astrophys.}

\def\physrep{Phys.~Rep.}

\def\jetp{Sov.~Phys.~JETP}
\def\njp{New J.~Phys.}
\def\pnas{Proc.~Nat.~Acad.~Sci.}
\def\ppcf{Plasma Phys.~Control.~Fusion}

\newcommand{\D}[2]{\frac{{\rm d} #2}{{\rm d} #1}}
\newcommand{\pD}[2]{\frac{\partial #2}{\partial #1}}
\newcommand\bb[1]{\mbox{\boldmath{$#1$}}}
\newcommand{\msb}[1]{\bb{\mathsf{#1}}}
\newcommand\grad{\bb{\nabla}}
\newcommand\bcdot{\,\bb{\cdot}\,}
\newcommand\btimes{\,\bb{\times}\,}
\newcommand{\const}{{\rm const}}
\newcommand{\rmd}{{\rm d}}
\newcommand{\kbcj}{k_{B\times J}}
\newcommand{\kbdj}{k_{B\cdot J}}

\begin{document}

\title{Tearing Instability and Current-Sheet Disruption in the Turbulent Dynamo}

\author{Alisa K.~Galishnikova}\email{alisag@princeton.edu}
\affiliation{Department of Astrophysical Sciences, Princeton University, 4 Ivy Lane, Princeton, NJ 08544, USA}
\author{Matthew~W.~Kunz}\email{mkunz@princeton.edu}%
\affiliation{Department of Astrophysical Sciences, Princeton University, 4 Ivy Lane, Princeton, NJ 08544, USA}
\affiliation{Princeton Plasma Physics Laboratory, PO Box 451, Princeton, NJ 08543, USA}
\author{Alexander A.~Schekochihin}
\affiliation{The Rudolf Peierls Centre for Theoretical Physics, University of Oxford, Clarendon Laboratory, Parks Road, Oxford, OX1 3PU, UK}
\affiliation{Merton College, Oxford OX1 4JD, UK}

\date{\today}

\begin{abstract}
Turbulence in a conducting plasma can amplify seed magnetic fields in what is known as the turbulent, or small-scale, dynamo. The associated growth rate and emergent magnetic-field geometry depend sensitively on the material properties of the plasma, in particular on the Reynolds number ${\rm Re}$, the magnetic Reynolds number ${\rm Rm}$, and their ratio ${\rm Pm}\equiv{\rm Rm}/{\rm Re}$. For ${\rm Pm} > 1$, the amplified magnetic field is gradually arranged into a folded structure, with direction reversals at the resistive scale and field lines curved at the larger scale of the flow. As the mean magnetic energy grows to come into approximate equipartition with the fluid motions, this folded structure is thought to persist. Using analytical theory and high-resolution MHD simulations with the {\tt Athena++} code, we show that these magnetic folds become unstable to tearing during the nonlinear stage of the dynamo for ${\rm Rm}\gtrsim 10^4$ and ${\rm Re}\gtrsim 10^3$. An ${\rm Rm}$- and ${\rm Pm}$-dependent tearing scale, at and below which folds are disrupted, is predicted theoretically and found to match well the characteristic field-reversal scale measured in the simulations. The disruption of folds by tearing increases the ratio of viscous-to-resistive dissipation. In the saturated state, the magnetic-energy spectrum exhibits a sub-tearing-scale steepening to a slope consistent with that predicted for tearing-mediated Alfv\'{e}nic turbulence. Its spectral peak appears to be independent of the resistive scale and comparable to the driving scale of the flow, while the magnetic energy resides in a broad range of scales extending down to the field-reversal scale set by tearing. Emergence of a degree of large-scale magnetic coherence in the saturated state of the turbulent dynamo may be consistent with observations of magnetic-field fluctuations in galaxy clusters and recent laboratory experiments.
\end{abstract}

\maketitle

\section{Introduction}\label{sec:intro}

Dynamo action refers to the amplification and subsequent maintenance of magnetic fields through the conversion of kinetic energy to magnetic energy~\cite{bs05,rincon19}. Reconnection refers to the annihilation and topological rearrangement of magnetic fields through the conversion of magnetic energy to kinetic energy~\cite{zy09,lu16}. Given this reciprocal relationship, it is somewhat surprising that studies of reconnection in the context of the dynamo are in their infancy. Indeed, with the exceptions of unpublished numerical work by Iskakov \& Schekochihin (2008) and Beresnyak (2012) (summarized in Ref.~\cite{Scheko_review}), there has been no systematic investigation of how reconnection affects the geometry of magnetic fields produced by the turbulent dynamo \footnote{Speculative arguments linking reconnection to the turbulent dynamo have been made in the past. Kulsrud \& Anderson~\cite{ka92} invoked Petschek's reconnection rate when modifying their kinematic dynamo theory to account for enhanced diffusion of small-scale magnetic fields, remarking presciently that ``As the field becomes more and more tangled, there will be places where the field is sharply reversed, and magnetic reconnection may set in, removing the sharpest kinks.'' A similar argument was made in Ref.~\cite{XuLazarian_2016} in the context of turbulent ``reconnection diffusion'' supplanting the usual microscopic resistive diffusion of dynamo-generated magnetic fields. Other models of dynamo invoking reconnection have either focused on the nature and statistics of magnetic dissipation~\cite{baggaley09} or treated reconnection as a means of reducing the back-reaction of the Lorentz force on the turbulent stretching motions~\cite{blackman96}.}.

At least part of the blame for the lack of progress on this front may be  attributed to the steep computational cost involved. It is now well established that very large Lundquist numbers are required for current sheets to undergo fast reconnection via the plasmoid (tearing) instability~\cite{loureiro07,lu16}, and large Lundquist numbers require high numerical resolution (the Lundquist number is the ratio of a sheet's resistive diffusion time to its Alfv\'{e}n-crossing time, and increases with decreasing resistivity). The fact that the turbulent dynamo is an inherently three-dimensional process~\cite{zeldovich57} further compounds the cost. But in the face of such computational adversity, theory can flourish, and one may take inspiration from recent analytical developments concerning the impact of resistive tearing modes on critically balanced, dynamically aligned magnetohydrodynamic (MHD) turbulence in the presence of a mean magnetic field~\cite{mallet17,lb17,bl17,lb20}. Recent simulations in reduced geometries appear to support those authors' conjecture that the three-dimensionally anisotropic Alfv\'{e}nic fluctuations that occur at the small scales of a turbulent cascade produce current sheets that are susceptible to disruption (either incomplete or complete) via tearing~\cite{walker18,dong18}.

Following this line of reasoning, here we use analytical arguments and high-resolution, visco-resistive MHD simulations to determine under what conditions the geometry of the magnetic field produced by the turbulent (or ``fluctuation'') dynamo is affected by the tearing instability. The idea itself has three simple ingredients~\cite{Scheko_review}. 

\begin{figure}
    \centering
    \includegraphics[width=\columnwidth]{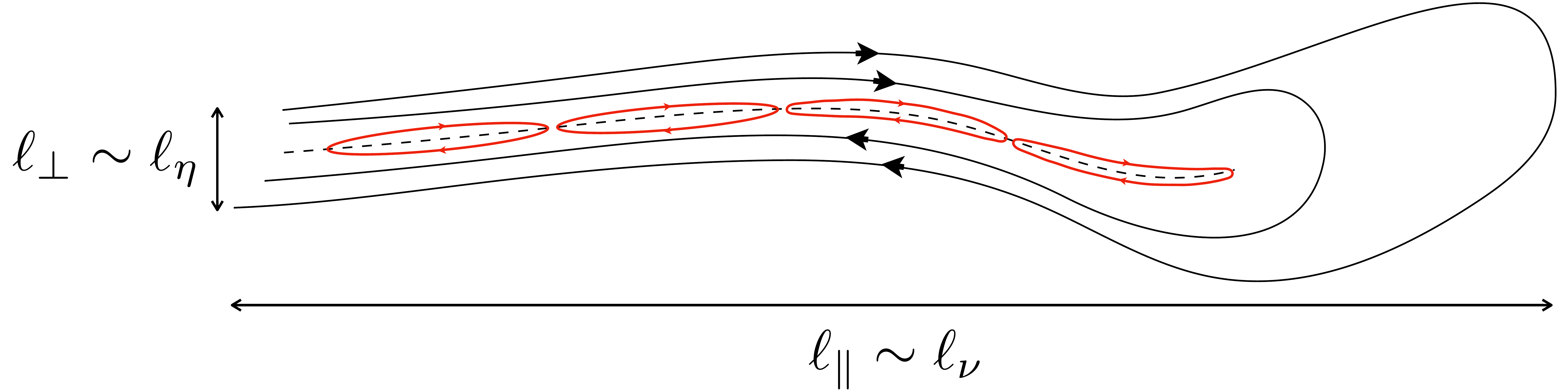}    \caption{An illustrative sketch of a characteristic magnetic fold produced during the kinematic stage of the turbulent dynamo, with its length $\ell_\parallel$ set by the viscous scale $\ell_\nu$ and its width $\ell_\perp$ set by the resistive scale $\ell_\eta$. The possibility that such a fold might be disrupted by resistive tearing modes (symbolized by the red islands), in either the kinematic or the nonlinear stage of the dynamo, is investigated in this paper.}
    \label{fig:sketch}
\end{figure}

First, a generic outcome of (the kinematic stage of) the turbulent dynamo is a dynamically important magnetic field exhibiting a folded geometry~\cite{Scheko_theory2,scheko02_sss}, with existing simulations predicting a characteristic length of the folds set by the viscous scale of the fluid motions, $\ell_\nu$, and a characteristic width of the folds related to the resistive scale, $\ell_\eta$~\cite{Scheko04sim,maron04}. When the ratio of the kinematic viscosity $\nu$ to the resistivity $\eta$ is large, i.e., when the magnetic Prandtl number ${\rm Pm}\equiv\nu/\eta\gg{1}$, these lengthscales satisfy $\ell_\nu \gg \ell_\eta$, and the folds may be viewed as elongated current sheets whose characteristic direction-reversing scale is much smaller than their characteristic coherence length (see Figure~\ref{fig:sketch}). 

This view of the turbulent dynamo as an efficient generator of thin current sheets motivates the second ingredient, namely, the tearing instability. Depending on the aspect ratio of the current sheet and the Lundquist number, the tearing instability triggers the onset of reconnection by perturbing the reconnecting field to spawn one, two, or even a whole chain of magnetic islands~\cite{fkr63,coppi76,loureiro07}. If allowed to proceed beyond its linear stage of growth, tearing undergoes nonlinear evolution that ultimately leads to the collapse and break-up of the current sheet~\cite{waelbroeck93,loureiro05,samtaney09,bhattacharjee09,uzdensky10,loureiro12}.

The third and final ingredient is an appreciation for the important role played in all this by the material properties of the plasma. Namely, the larger the value of ${\rm Pm}$, the more spatially anisotropic the dynamo-generated folds become, with the length-to-width ratio increasing approximately as ${\rm Pm}^{1/2}$ in the kinematic regime~\cite{Scheko04sim}. This arrangement is particularly conducive to tearing, as it allows for modes with larger values of the tearing-instability parameter, $\Delta'$. On the other hand, large values of ${\rm Pm}$ slow down tearing modes through viscous damping of the fluid motions~\cite{porcelli87}. For example, the critical Lundquist number for plasmoid instability in a ${\rm Pm}\gg{1}$ Sweet--Parker current sheet {\em increases} with ${\rm Pm}$, making it more difficult to trigger fast reconnection~\cite{loureiro13}. Clearly, there is a sweet spot in the values of ${\rm Pm}$.

This qualitative argument is made more quantitative in \S\ref{sec:expectations}, where we develop a theory to estimate a ``tearing scale'' at and below which tearing disrupts the dynamo-generated folds. This theory does not, however, address how this physics translates into the fluctuation statistics and spectra, the ratio of viscous to resistive heating, and the eventual structure of the dynamo-generated magnetic field in its saturated state. To find that out, we perform a series of numerical simulations, which are described and analyzed in \S\ref{sec:sim}. We close in \S\ref{sec:summary} with some thoughts on how this tearing-mediated dynamo might manifest in astrophysical systems, and what its implications are for the production of large-scale fields in turbulent astrophysical plasmas. A related finding of our analysis is that the peak of the magnetic-energy spectrum in the saturated state of the dynamo appears to occur at large scales and to be independent of the resistivity when the latter is sufficiently small. In other words, the saturated state of the small-scale dynamo is characterized by a degree of large-scale coherence in the amplified magnetic field.

\section{Theoretical considerations}\label{sec:expectations}

In this Section, we summarize the salient features of the ${\rm Pm}\ge{1}$ fluctuation dynamo, as suggested by theory and evidenced by low- and intermediate-resolution numerical simulations (\S\ref{sec:folds}), and of the theory of the resistive tearing instability (\S\ref{sec:tearing}). These features are then combined in \S\ref{sec:theory} to obtain a theory for how tearing might disrupt dynamo-generated magnetic folds.

\subsection{Generation and persistence of magnetic folds}\label{sec:folds}

Consider a statistically homogeneous MHD plasma with constant magnetic resistivity $\eta$ and kinematic viscosity~$\nu$ (${\ge}\eta$), in which an initially weak, zero-net-flux magnetic field is amplified via random stretching by three-dimensional, incompressible turbulence. We take this turbulence to consist of fluid motions that are injected with root-mean-square (rms) velocity $U$ at the outer (forcing) scale $L$ and cascaded conservatively through an inertial range down to a viscous scale $\ell_\nu$, at and below which dissipation occurs. This assumes that the magnetic field is weak enough that the Lorentz force is negligible throughout this inertial range --- the so-called ``kinematic'' stage. For a Kolmogorov cascade~\cite{Kolmogorov1941}, the typical velocity increment at scale $\ell$ is given by $\delta u_\ell \sim U (\ell/L)^{1/3}$, resulting in a nonlinear eddy turnover time $\tau_{\rm nl}$ that progressively decreases at smaller and smaller scales, {\em viz.}, 
\begin{equation}
    \tau^{-1}_{\rm nl} \sim \frac{\delta u_\ell}{\ell} \sim \frac{U}{L} \left(\frac{\ell}{L}\right)^{-2/3}.
\end{equation}
In this case, the maximal stretching rate, 
\begin{equation}\label{eqn:ROSkinematic}
    \tau^{-1}_{\rm nl,min} \sim \frac{U}{L}\,{\rm Re}^{1/2} \quad\textrm{(kinematic stage),}
\end{equation}
occurs at the viscous scale $\ell_\nu \sim L\,{\rm Re}^{-3/4}$, where ${\rm Re}\doteq UL/\nu\ge 1$ is the Reynolds number. When the magnetic field is too weak to exert any influence on these viscous-scale motions, the rms magnetic-field strength $B_{\rm rms}$ (defined throughout the paper in velocity units assuming a constant background density) grows exponentially at this ${\rm Re}$-dependent rate, 
\begin{equation}
    \D{t}{\ln B_{\rm rms}} \sim \frac{U}{L}\, {\rm Re}^{1/2}\quad\textrm{(kinematic stage)}.
\end{equation}

During this exponential amplification, the magnetic field is stretched, folded, and ultimately organized into a highly intermittent patchwork of long, thin structures whose energy spectrum $M(k) \propto k^{3/2}$ (here $k$ is the wavenumber) peaks at the smallest available scale on which the magnetic field can reverse its direction, {\it viz.}, the resistive scale $\ell_\eta$~\cite{Kazantsev,ka92,scheko02}. In this folded geometry, there is an anti-correlation between the field-line curvature and the field strength~\cite{Scheko_theory2,scheko02_sss,Scheko04sim}: the magnetic field is weakest in the regions of curved field (bends), and strongest in the regions where it reverses its direction. To obtain an estimate for $\ell_\eta$, we balance the maximal (viscous-scale) nonlinear stretching rate, $\tau^{-1}_{\rm nl,min} \sim (U/L)\,{\rm Re}^{1/2}$, with the rate of resistive decay of the folds, $\tau^{-1}_\eta \sim \eta/\ell^2_\eta$, finding $\ell_\eta \sim L\,{\rm Re}^{-1/4}{\rm Rm}^{-1/2}$, where ${\rm Rm}\equiv UL/\eta$ is the magnetic Reynolds number. The ratio of the viscous to resistive scales then satisfies $\ell_\nu/\ell_\eta \sim ({\rm Rm}/{\rm Re})^{1/2} \equiv {\rm Pm}^{1/2}$. Ref.~\cite{Scheko_theory2} showed that the characteristic parallel length of the magnetic folds, $\ell_\parallel$, is inherited from the velocity fluctuations with the fastest rate of strain, in which case $\ell_\parallel \sim \ell_\nu$ during the kinematic stage. Thus, the value of ${\rm Pm}$ controls the aspect ratio of the folds, with ${\rm Pm}\gg{1}$ implying large-aspect-ratio current sheets. In \S\ref{sec:theory} we show that, despite these potentially large aspect ratios, the lifetime of the folds during the kinematic stage is too short for the tearing instability to grow effectively. 

Once the mean magnetic energy becomes comparable to the energy of the viscous-scale motions ({\em viz.}, $B^2_{\rm rms} \sim U^2\,{\rm Re}^{-1/2}$, the kinematic stage ends and the dynamo becomes nonlinear. Namely, the Lorentz force due to the spatially coherent magnetic folds back-reacts on the viscous-scale motions and suppresses their ability to amplify the magnetic field~\cite{cattaneo96,kim99,boldyrev01,scheko04a,Scheko04sim,ct09,seta20,St-Onge2020}. As a result, progressively larger (and slower) eddies are responsible for amplifying the field, while the eddies whose energies are lower, $\delta u^2_\ell \sim U^2(\ell/L)^{2/3} \lesssim B^2_{\rm rms}$, are suppressed. This leads to some steepening of the kinetic-energy spectrum just below the energy-equipartition scale. The result is a resistive ``selective decay'' of the magnetic energy at scales too small to be sustainable by the weakened stretching~\cite{Scheko02theory,Scheko04sim,maron04}. With the maximal stretching rate now being given by
\begin{equation}\label{eqn:ROSnonlinear}
    \tau^{-1}_{\rm nl,min} \sim \frac{U}{L} \left(\frac{U}{B_{\rm rms}}\right)^2 \quad\textrm{(nonlinear stage),}
\end{equation}
the magnetic energy grows secularly, with
\begin{equation}
    \D{t}{B^2_{\rm rms}} \sim \frac{U^3}{L} = \const \quad\textrm{(nonlinear stage)}
\end{equation}
implying $B_{\rm rms} \propto t^{1/2}$~\cite{Scheko02theory,cho09,beresnyak12}. 

During this stage of secular growth, the resistive scale that is obtained by balancing stretching with resistive decay satisfies $\ell_\eta \sim L\,(B_{\rm rms}/U) {\rm Rm}^{-1/2} \propto t^{1/2}$, and thus the energy-containing scale of the magnetic field shifts gradually towards larger scales. The length of the folds increases as well, matching that of the maximally stretching eddies, {\it viz.}, $\ell_\parallel \sim L\,(B_{\rm rms}/U)^3 \propto t^{3/2}$. Accordingly, the folds become further elongated: $\ell_\parallel/\ell_\eta \propto t$. In \S\ref{sec:theory}, we show that it is during this stage that the dynamo-generated current sheets first become susceptible to tearing, thereby modifying these scalings.

Eventually, the magnetic energy reaches approximate equipartition with the kinetic energy at the outer scale, $B^2_{\rm rms} \sim U^2$. In this saturated state, all but the largest eddies are suppressed by the magnetic tension associated with the dynamically important, folded magnetic field. In low- and intermediate-resolution simulations of the ${\rm Pm}\ge 1$ dynamo~\cite{maron04,Scheko04sim}, this folded geometry is found to persist in the saturated state, with a parallel coherence length set by the outer-scale motions and a perpendicular field-reversal scale that remains proportional to the resistive scale. With the latter being determined by a balance between the stretching rate of the outer-scale eddies, $\tau^{-1}_{\rm nl} \sim U/L$, and $\tau^{-1}_\eta \sim \eta/\ell^2_\eta$, we obtain $\ell_\eta \sim L\, {\rm Rm}^{-1/2}$, from which an aspect ratio $\ell_\parallel/\ell_\eta \sim {\rm Rm}^{1/2}$ follows. In \S\ref{sec:theory}, we argue that these scalings should fail at large Rm due to  disruption of the folds by tearing instability. Before doing so, we recapitulate briefly the theory of tearing instability in super-critical current sheets.

\subsection{Tearing modes in super-critical current sheets}\label{sec:tearing}

A current sheet with length $\ell$ and characteristic half-thickness $\lambda$ is deemed ``super-critical'' if there is a tearing-mode wavenumber $k_{\rm t}\gtrsim\ell^{-1}$ for which the stability parameter $\Delta'=\Delta'(k_{\rm t})>0$. For a Harris-sheet (i.e., tanh) profile, $\Delta' \lambda = 2(1/k_{\rm t}\lambda - k_{\rm t}\lambda)$, so that modes with $\lambda/\ell\lesssim k_{\rm t}\lambda \ll 1$ are the most susceptible to tearing; for a sinusoidal profile,  $\Delta' \lambda \simeq (8/\pi)(k_{\rm t}\lambda)^{-2}$ when $k_{\rm t}\lambda\ll 1$~\cite{bl18}. In either case, for tearing modes with $\Delta' \lambda \ll (S_\lambda k_{\rm t}\lambda)^{1/3}$, where $S_\lambda\equiv \lambda v_{\rm A,\lambda}/\eta$ is the Lundquist number of the sheet and $v_{\rm A,\lambda}$ is the Alfv\'{e}n speed of the reversing field (the so-called ``FKR regime''~\cite{fkr63}), the growth is exponential at the rate $\gamma_{\rm FKR} \sim (v_{\rm A,\lambda}/\lambda) S^{-3/5}_\lambda (k_{\rm t}\lambda)^{2/5} (\Delta' \lambda)^{4/5}$. For a current-sheet profile satisfying $\Delta' \lambda \sim (k_{\rm t}\lambda)^{-n}$ at $k_{\rm t}\lambda \ll 1$, the growth rate $\gamma_{\rm FKR}$ is largest at the smallest available wavenumber ($k_{\rm t}\ell\sim 1$) for $n>1/2$. In this case, low-aspect-ratio sheets will develop tearing perturbations comprising just one or two islands. If instead the current sheet is proportionally thin enough that $\Delta' \lambda \gtrsim (S_\lambda k_{\rm t}\lambda)^{1/3}$, i.e., if $\lambda/\ell \lesssim k_{\rm t}\lambda\lesssim S^{-1/(3n+1)}_\lambda$, the growth rate becomes independent of $\Delta'$---the so-called ``Coppi regime''~\cite{coppi76}, with $\gamma_{\rm Coppi} \sim (v_{\rm A,\lambda}/\lambda) S^{-1/3}_\lambda (k_{\rm t}\lambda)^{2/3}$. In this case the growth rate {\em increases} with increasing $k_{\rm t}$, signaling that high-aspect-ratio sheets will spawn whole chains comprising ${\sim}k_{\rm t}\ell\gg 1$ islands. One can then show by balancing the FKR and Coppi rates~\cite{UL16} that, if the Coppi regime is accessible, the maximally growing tearing mode has $\gamma_{\rm t,max} \sim (v_{\rm A,\lambda}/\lambda) S^{-(n+1)/(3n+1)}_\lambda$ at $k_{\rm t,max}\lambda \sim S^{-1/(3n+1)}_\lambda$.

The above tearing scalings are valid only at ${{\rm Pm}\lesssim 1}$. For ${{\rm Pm}\gg 1}$, they must be modified to account for viscous suppression of the fluid motions involved in the tearing modes~\cite{porcelli87}. In this case, the corresponding ``FKR'' and ``Coppi'' growth rates are given by $\gamma_{\rm FKR}\sim (v_{\rm A,\lambda}/\lambda)S^{-2/3}_\lambda {\rm Pm}^{-1/6} (k_{\rm t}\lambda)^{1/3} (\Delta'\lambda)$ and $\gamma_{\rm Coppi} \sim (v_{\rm A,\lambda}/\lambda) S^{-1/3}_\lambda {\rm Pm}^{-1/3}(k_{\rm t}\lambda)^{2/3}$. For $\Delta'\lambda \sim (k_{\rm t}\lambda)^{-n}$ with $n>1/3$, there is again a maximally growing tearing mode intermediate between between these two regimes, which may be obtained as before by balancing the FKR and Coppi rates. The result is
\begin{subequations}\label{eqn:maxtearing}
\begin{align}
    \frac{\gamma_{\rm t,max} \lambda}{v_{\rm A,\lambda}} &\sim S^{-(n+1)/(3n+1)}_\lambda {\rm Pm}^{-n/(3n+1)} ,\label{eqn:gmax}\\*
    k_{\rm t,max}\lambda &\sim S^{-1/(3n+1)}_\lambda {\rm Pm}^{1/2(3n+1)} ,\label{eqn:kmax}
\end{align}
\end{subequations}
provided that $k_{\rm t,max}\ell\gtrsim 1$~\cite{Scheko_review}.

\subsection{Tearing meets dynamo}\label{sec:theory}

For tearing to be relevant during the fluctuation dynamo, the maximum growth rate for tearing of a current sheet must be larger than both the sheet's resistive decay rate and its decorrelation rate, {\em viz.}, $\gamma_{\rm t,max}\tau_\eta\gtrsim 1$ and $\gamma_{\rm t,max}\tau_{\rm nl,min} \gtrsim 1$. These requirements are tantamount to asking whether the maximum current-sheet thickness at which tearing can onset (denoted by $\lambda_\ast$) is larger than the resistive scale, $\ell_\eta \sim (\eta\tau_{\rm nl,min})^{1/2}$. Here we use Eq.~\eqref{eqn:maxtearing} to determine $\lambda_\ast$ via the condition $\gamma_{\rm t,max}\tau_{\rm nl,min}\sim 1$ in each stage of the dynamo and ask whether it is ${\gtrsim}\ell_\eta$; we also check that $k_{\rm t,max}\ell_\parallel\gtrsim 1$ (i.e., that the Coppi regime is accessible). When doing so, we associate $v_{\rm A,\lambda}$ in Eq.~\eqref{eqn:maxtearing} with the strength of the local dynamo-generated field whose reversal scale is $\lambda = \lambda_\ast$.

As explained in \S\ref{sec:folds}, during the kinematic stage, $\tau_{\rm nl,min}^{-1} \sim (U/L) {\rm Re}^{1/2}$ [see Eq.~\eqref{eqn:ROSkinematic}] and so $\ell_\eta \sim L\,{\rm Rm}^{-3/4}{\rm Pm}^{1/4}$. Demanding that $\lambda_\ast\gtrsim\ell_\eta$ is then equivalent to demanding that $B^2_{\rm rms} \gtrsim U^2\, {\rm Re}^{-1/2}$. This is the same as the energy of the viscous-scale eddies, so such field strengths are greater than those attained during the kinematic stage. In other words, {\em all current sheets produced during the kinematic stage should diffuse resistively before tearing can onset} (this statement is independent of $n$). By the end of the kinematic stage, however, $\lambda_\ast\sim\ell_\eta$. For tearing to onset during the subsequent nonlinear (secular-growth) stage, $\lambda_\ast$ must then grow in time faster than $\ell_\eta \sim L\, (B_{\rm rms}/U){\rm Rm}^{-1/2}\propto t^{1/2}$. We now show that this is indeed the case.

Using Eqs~\eqref{eqn:ROSnonlinear} and \eqref{eqn:gmax} and comparing $\gamma_{\rm t,max}$ to $\tau^{-1}_{\rm nl,min} \sim (U/L)(U/B_{\rm rms})^2\propto t^{-1}$ during the nonlinear stage, we find that 
\begin{equation}\label{eqn:amax_L}
    \frac{\lambda_\ast}{L} \sim \left[ \dfrac{ (B^2_{\rm rms}/U^2)^{4n+1} }{{\rm Rm}^{n+1} {\rm Pm}^{n}} \right]^{1/2(2n+1)} \propto t^{\frac{4n+1}{4n+2}} ,
\end{equation}
and so
\begin{equation}\label{eqn:amax_ln}
    \frac{\lambda_\ast}{\ell_\eta} \sim \left[ \frac{B^2_{\rm rms}}{U^2} {\rm Re}^{1/2} \right]^{n/(2n+1)} \propto t^{n/(2n+1)}.
\end{equation}
Although $\ell_\eta$ increases in time, $\lambda_\ast$ does so even faster, affording the possibility of tearing disrupting the current sheets before they  diffuse resistively. With $k_{\rm t,max}$ given by Eq.~\eqref{eqn:kmax} and $\ell_\parallel \sim L\,(B_{\rm rms}/U)^3$ (see \S\ref{sec:folds}), a typical current sheet should then spawn
\begin{equation}\label{eqn:ktmax}
    N \sim k_{\rm t,max}\ell_\parallel \sim \left[ \frac{B^2_{\rm rms}}{U^2} {\rm Re}^{1/2} \right]^{n/(2n+1)} {\rm Pm}^{1/2} \gtrsim 1
\end{equation}
magnetic islands. Thus, unless ${\rm Re}\lesssim 1$, {\em a nonlinear dynamo is a tearing-limited dynamo}.

Eventually, the dynamo should saturate with a near-equipartition magnetic field, $B_{\rm rms}\sim U$, so that, using Eqs.~\eqref{eqn:amax_L} and \eqref{eqn:amax_ln},
\begin{subequations}\label{eqn:tearingscale}
\begin{align}
    \lambda_\ast &\sim L \, \Bigl( {\rm Rm}^{n+1} {\rm Pm}^{n} \Bigr)^{-1/2(2n+1)}  \label{eqn:amax_sat}\\*
    \mbox{} &\sim \ell_\eta \, {\rm Re}^{n/2(2n+1)} .\label{eqn:amax_sat2}
\end{align}
\end{subequations}
Thus, for there to be a range of scales on which tearing acts much faster than resistive decay and nonlinear decorrelation (i.e., $\lambda_\ast\gg \ell_\eta$), we require that
\begin{equation}\label{eqn:Recrit}
    {\rm Re}^{n/2(2n+1)}\gg 1.
\end{equation}
For $n=1$, this gives ${\rm Re}^{1/6}\gg 1$; for $n=2$, ${\rm Re}^{1/5}\gg 1$ \footnote{The requirement \eqref{eqn:Recrit} that ${\rm Re}$ be large enough for there to be a range of scales between $\lambda_\ast$ and $\ell_\eta$ is much more forgiving (in terms of computational expense) than the corresponding requirement governing the possible disruption by tearing of  strong Alfv\'{e}nic turbulence in the presence of a guide field. In the latter case, assuming a critically balanced, dynamically aligned cascade~\cite{boldyrev06,chandran15,ms17}, one requires ${\rm Rm}^{2n/3(4n+3)} (1+{\rm Pm})^{-2(7n+3)/3(4n+3)}\gg 1$~\cite{Scheko_review}. For $n=1$, this gives ${\rm Rm}^{2/21}(1+{\rm Pm})^{-20/21}\gg 1$; for $n=2$, ${\rm Rm}^{4/33}(1+{\rm Pm})^{-34/33}\gg 1$. In either case, much smaller diffusivities (and thus higher numerical resolutions) are required for tearing disruption of an Alfv\'{e}nic cascade than for tearing disruption of dynamo-produced folds. The physical reason is that a long inertial range is needed for the three-dimensionally anisotropic eddies in an Alfv\'{e}nic cascade to realize current sheets that are thin enough, and have large enough aspect ratios in the plane perpendicular to the guide field, to be supercritical. By contrast, elongated folds are produced naturally by the turbulent dynamo when ${\rm Pm}\gg 1$.}. If this inequality is satisfied, then in saturation we have from Eq.~\eqref{eqn:amax_sat} that
\begin{equation}\label{eqn:lambdastar}
    \lambda_\ast \sim 
    \begin{cases}
        ~L\, {\rm Rm}^{-1/3} {\rm Pm}^{-1/6} & \textrm{for }n=1 ;\\
        ~L\, {\rm Rm}^{-3/10} {\rm Pm}^{-1/5} & \textrm{for }n=2 .
    \end{cases}
\end{equation}
These predictions, alongside those that tearing should not operate during the kinematic stage and that the magnetic field should exhibit a reversal scale during the nonlinear stage that evolves as $\lambda_\ast\propto t^{5/6}$ ($n=1$) or $t^{9/10}$ ($n=2$) [see Eq.~\eqref{eqn:amax_L}], rather than as $\ell_\eta \propto t^{1/2}$, are tested by the numerical simulations in \S\ref{sec:sim}.

If ${\rm Re}$ is sufficiently large that $\lambda_\ast$ is well separated from~$\ell_\eta$ [see Eq.~\eqref{eqn:amax_sat2}], then there is the additional question of how the magnetic field is arranged at sub-$\lambda_\ast$ scales, e.g., what is its energy spectrum. This depends on whether the tearing of the folds proceeds long enough to go nonlinear, induce current-sheet collapse, and perhaps trigger the onset of plasmoid-dominated reconnection. Because $k_{\rm t,max}\ell_\parallel \gtrsim 1$ in saturation [see Eq.~\eqref{eqn:ktmax}], the Coppi regime is accessible  and so we do not anticipate a nonlinear ``Rutherford'' stage of secular growth of the island width~\cite{Rutherford73}. (Put differently, the island widths are already comparable to the inner-layer width $\delta_{\rm in}$ of the tearing current sheet at the end of the linear growth of tearing, and so no further evolution is needed to obtain $\Delta'\delta_{\rm in}\sim 1$.) The X-point(s) formed by the tearing mode should then rapidly collapse into thin secondary sheets, resulting in Sweet--Parker-like growth of the reconnected flux on a time scale comparable to $\gamma^{-1}_{\rm t,max}$~\cite{loureiro05}. An outstanding question then is whether plasmoid-dominated, steady-state, fast reconnection can onset (e.g., see Section 12.4.4 of Ref.~\cite{Scheko_review}), but that is unlikely to be verifiable by numerical simulations at resolutions that are currently feasible.

In what follows, we {\em assume} that steady-state, fast reconnection is not occurring, and that the role of tearing is simply to break up the folds into a succession of smaller structures (such an assumption is supported by the numerical results in \S\ref{sec:sim}). We then argue that the turbulence at scales below $\lambda_\ast$ should be similar to the tearing-mediated turbulence proposed in Refs~\cite{lb17,mallet17}, with the only difference being that the direction of the ``local mean field'' is that of the strongly fluctuating dynamo field at larger scales. If this is true, then {\em the energy spectrum of the cascade should exhibit a $k^{-11/5}~(n=1)$ or $k^{-19/9}~(n=2)$ spectral envelope at scales below $\lambda_\ast$}~\cite{lb17,mallet17}. Both of these slopes are steeper than Kolmogorov. The derivation of this spectrum rests on the assumption that the turbulence at sub-$\lambda_\ast$ scales is approximately Alfv\'{e}nic and critically balanced. For each scale $\lambda\lesssim\lambda_\ast$, the nonlinear turnover time $\tau_{\rm nl}$ associated with the Alfv\'{e}nic motions on that scale is comparable to the linear tearing timescale $\gamma^{-1}_{\rm t,max}$ at the same scale. Assuming that tearing leads to a negligible amount of dissipation in the tearing-mediated cascade (that is, tearing only determines the lifetime and structure of the sub-$\lambda_\ast$ fluctuations), Eq.~\eqref{eqn:gmax} with $\gamma_{\rm t,max}v^2_{\rm A,\lambda}\sim \const$ below $\lambda_\ast$ leads to the scaling $v_{\rm A,\lambda}\propto \lambda^{(2n+1)/(4n+1)}$, from which the aforementioned spectra follow.

In the next section, we test these predictions with high-resolution numerical simulations across a wide range of ${\rm Rm}$ and ${\rm Pm}$.

\section{Numerical results}\label{sec:sim}

\subsection{Method of solution and dimensionless free parameters}

% TABLE 1

\begin{table}
  \begin{center}
  \begin{tabular}{|l|cccccccc|}
     \hline
      $N^3$ & run & $10^7\eta$ & {\rm Pm} & ${\rm Re}_{\rm sat}$ & ${\rm Rm}_{\rm sat}$ & $u_{\rm rms,kin}$ & $u_{\rm rms,sat}$ & $B_{\rm rms,sat}$ \\[3pt]
     \hline
     \hline
     \parbox[t]{2mm}{\multirow{7}{*}{\rotatebox[origin=c]{90}{280$^3$}}} & a1 & 200 &1 & 920 & 920 & 0.14 & 0.12 & 0.06\\
     & a2 & 200 &10 & 76 & 760 & 0.11 & 0.10 & 0.07\\
     & a3 & 200 &50 & 12.9 & 640 & 0.11 & 0.08 & 0.08\\
     & a4 & 200 &100 & 7.0 & 700 & 0.14 & 0.09 & 0.10\\
     & a5 & 200 &200 & 3.6 & 720 & 0.12 & 0.09 & 0.11\\
     & a6 & 200 &300 & 2.5 & 750 & 0.12 & 0.09 & 0.12\\
     & a7 & 200 &500 & 1.45 & 720 & 0.13 & 0.09 & 0.12\\
     \hline
     \parbox[t]{2mm}{\multirow{7}{*}{\rotatebox[origin=c]{90}{560$^3$}}} & b1 & 100 &1 & 2200 & 2200 & 0.18 & 0.14 & 0.09\\
     & b2 & 100 &10 & 189 & 1890 & 0.18 & 0.12 & 0.09\\
     & b3 & 100 &50 & 33 & 1640 & 0.19 & 0.10 & 0.11\\
     & b4 & 100 &100 & 18.7 & 1870 & 0.19 & 0.12 & 0.13\\
     & b5 & 100 &200 & 10.2 & 2000 & 0.21 & 0.13 & 0.16\\
     & b6 & 100 &300 & 7.2 & 2200 & 0.22 & 0.14 & 0.18\\
     & b7 & 100 &500 & 4.7 & 2300 & 0.22 & 0.15 & 0.19\\
     \hline
     \parbox[t]{2mm}{\multirow{7}{*}{\rotatebox[origin=c]{90}{1120$^3$}}} & c1 & 25 &1 & 10400 & 10400 & 0.26 & 0.16 & 0.12\\
     & c2 & 25 &10 & 940 & 9400 & 0.26 & 0.15 & 0.13\\
     & c2$\star$& 25 &10 & 920 & 9200 & 0.17 & 0.15 & 0.13\\
     & b2$\diamond$ & 100 & 10 & 250 & 2500 & 0.18 & 0.16 & 0.12\\
     & c3 & 25 &50 & 163 & 8200 & 0.24 & 0.13 & 0.13\\
     & c4 & 25 &100 & 82 & 8200 & 0.23 & 0.13 & 0.14\\
     & c5 & 25 &200 & 47 & 9300 & 0.24 & 0.15 & 0.17\\
     & c6 & 25 &300 & 33 & 10000 & 0.30 & 0.16 & 0.19\\
     & c7 & 25 &500 & 22 & 11200 & 0.31 & 0.18 & 0.23\\
     \hline
     \parbox[t]{2mm}{\multirow{4}{*}{\rotatebox[origin=c]{90}{2240$^3$}}} & d1 & 6 &1  & 50000 & 50000 & 0.32 & 0.19 & 0.16\\
     & d2 & 6 &10 & 4800 & 48000 & 0.32 & 0.18 & 0.17\\
     & d2$\star$& 6 &10 & 5200 & 52000 & 0.25 & 0.20 & 0.16\\
     & c2$\diamond$ & 25 & 10 & 1130 & 11300 & 0.18 & 0.18 & 0.16\\
     & d3 & 6 &50 &  870 & 43000 & 0.32 & 0.16 & 0.17\\
     & d4 & 6 &100 & 430 & 43000 & 0.32 & 0.16 & 0.16\\
     \hline
  \end{tabular}
  \caption{Run parameters at different resolutions $N^3$. The subscripts ``sat'' and ``kin'' refer, respectively, to values measured during the saturated and kinematic stages. Versions of runs c2 and d2 using third-order Runge-Kutta time integration and third-order spatial reconstruction were also performed, here marked by a ``$\star$''. Versions of runs b2 and c2 with the same $\eta$ and ${\rm Pm}$ but performed at twice the resolution are marked by a ``$\diamond$''. Runs with ${\rm Pm}=1$ are slightly under-resolved during the kinematic stage.}  \label{tab:runs}
  \end{center}
\end{table}

We employ the {\tt Athena++} code framework~\cite{stone20} to solve the equations of non-relativistic, compressible magnetohydrodynamics (MHD) in conservative form. These are: the continuity equation,
\begin{equation}\label{eqn:cont}
    \pD{t}{\rho} + \grad\bcdot(\rho\bb{u}) = 0 ,
\end{equation}
the momentum equation,
\begin{equation}\label{eqn:mom}
    \pD{t}{\rho\bb{u}} + \grad\bcdot\biggl[ \rho\bb{u}\bb{u} - \bb{B}\bb{B} + \biggl(\rho C^2+\frac{B^2}{2}\biggr)\msb{I} - \rho\nu \msb{W} \biggr] = \rho\bb{f},
\end{equation}
and the induction equation,
\begin{equation}\label{eqn:ind}
    \pD{t}{\bb{B}} - \grad\btimes(\bb{u}\btimes\bb{B} - \eta\bb{J}) = 0 ,
\end{equation}
where $\rho$ is the mass density, $\bb{u}$ is the fluid velocity, $\bb{B}$ is the magnetic field, $\bb{J}=\grad\btimes\bb{B}$ is the current density,
\begin{equation}
    \msb{W} \equiv \grad\bb{u} + (\grad\bb{u})^\top - \frac{2}{3} (\grad\bcdot\bb{u})\msb{I}
\end{equation}
is the (traceless, symmetric) rate-of-strain tensor, and $\msb{I}$ is the unit dyadic. Equations~\eqref{eqn:mom} and \eqref{eqn:ind} include explicit momentum and magnetic diffusion with spatially uniform kinematic viscosity $\nu$ and Ohmic resistivity $\eta$. In writing equation~\eqref{eqn:mom}, we have adopted an isothermal equation of state with constant sound speed $C$ and included a driving term $\bb{f}$, specified below.

Equations~\eqref{eqn:cont}--\eqref{eqn:ind} are solved using the {\tt Athena++} code framework~\cite{stone20}. {\tt Athena++} is a widely used, finite-volume, astrophysical MHD code that uses a directionally unsplit Godunov scheme for MHD alongside constrained transport on a staggered grid to conserve the divergence-free property for magnetic fields, $\grad\bcdot\bb{B}=0$. All runs use the HLLD Riemann solver to calculate the fluxes. All but two of them employ a second-order-accurate van Leer integration algorithm with second-order-accurate piecewise linear spatial reconstruction; those exceptional two use third-order-accurate Runge-Kutta integration with third-order-accurate piecewise parabolic reconstruction. All these runs are summarized in Table~\ref{tab:runs}.

In each simulation, low-Mach-number turbulence with rms fluid velocity $u_{\rm rms}\approx 0.1C$ is driven in a three-dimensional, periodic box of size $L^3$ using an incompressible, zero-mean-helicity, random forcing $\bb{f}$. At each simulation time step, the Fourier coefficients $\bb{f}_{\!\bm{k}}$ are independently generated from a Gaussian-random field in the wavenumber range $k\in[1,2]k_0$, where $k_0\equiv 2\pi/L$ is the box wavenumber, and constrained to satisfy $\bb{k}\bcdot\bb{f}_{\!\bm{k}}=0$. The resulting force is then inverse-Fourier transformed, shifted to ensure no net momentum injection, and normalized to provide constant power per unit volume. The force is then time-correlated by an Ornstein--Uhlenbeck process with correlation time $t_{\rm corr,f} = 10 (k_0 C)^{-1} \approx (k_0 u_{\rm rms})^{-1}$. Across all of our simulations, the rms density fluctuation is never more than one percent.

We define the Reynolds number ${\rm Re} \equiv u_{\rm rms}/k_0\nu$ and the magnetic Reynolds number ${\rm Rm} \equiv u_{\rm rms}/k_0\eta$, for which ${\rm Pm}={\rm Rm}/{\rm Re} \ge 1$. We vary ${\rm Re}$ and ${\rm Rm}$ across the suite of simulations using grid resolutions $280^3$, $560^3$, $1120^3$, and $2240^3$. For each resolution, we set $\eta$ and then vary $\nu$ while keeping ${\rm Pm}\geq 1$. In doing so, we explore a wide range of plasma parameters, with ${\rm Rm} \sim 10^3$--$5\times 10^4$ and ${\rm Pm} \in [1,500]$. Two additional simulations (runs b2$\diamond$ and c2$\diamond$) were performed to test the robustness of our results by doubling the resolution at fixed $\nu$ and $\eta$. All simulations are initialized with density $\rho = \rho_0 = \const$, fluid velocity $\bb{u}=0$, and zero-net-flux magnetic field $\bb{B}$ with energy randomly distributed at wavenumbers $k\in[1,2]k_0$ and rms field strength $B_{\rm rms}$ such that $\beta_0 \equiv 2 \rho_0 C^2/B^2_{\rm rms} = 5\times10^5$. Henceforth, all quantities are normalized so that $L=C=\rho_0=1$.

\subsection{Qualitative evolution: evidence of magnetic folds and their disruption at large ${\rm Rm}$}

\begin{figure*}
    \centering
    \includegraphics[width=\textwidth]{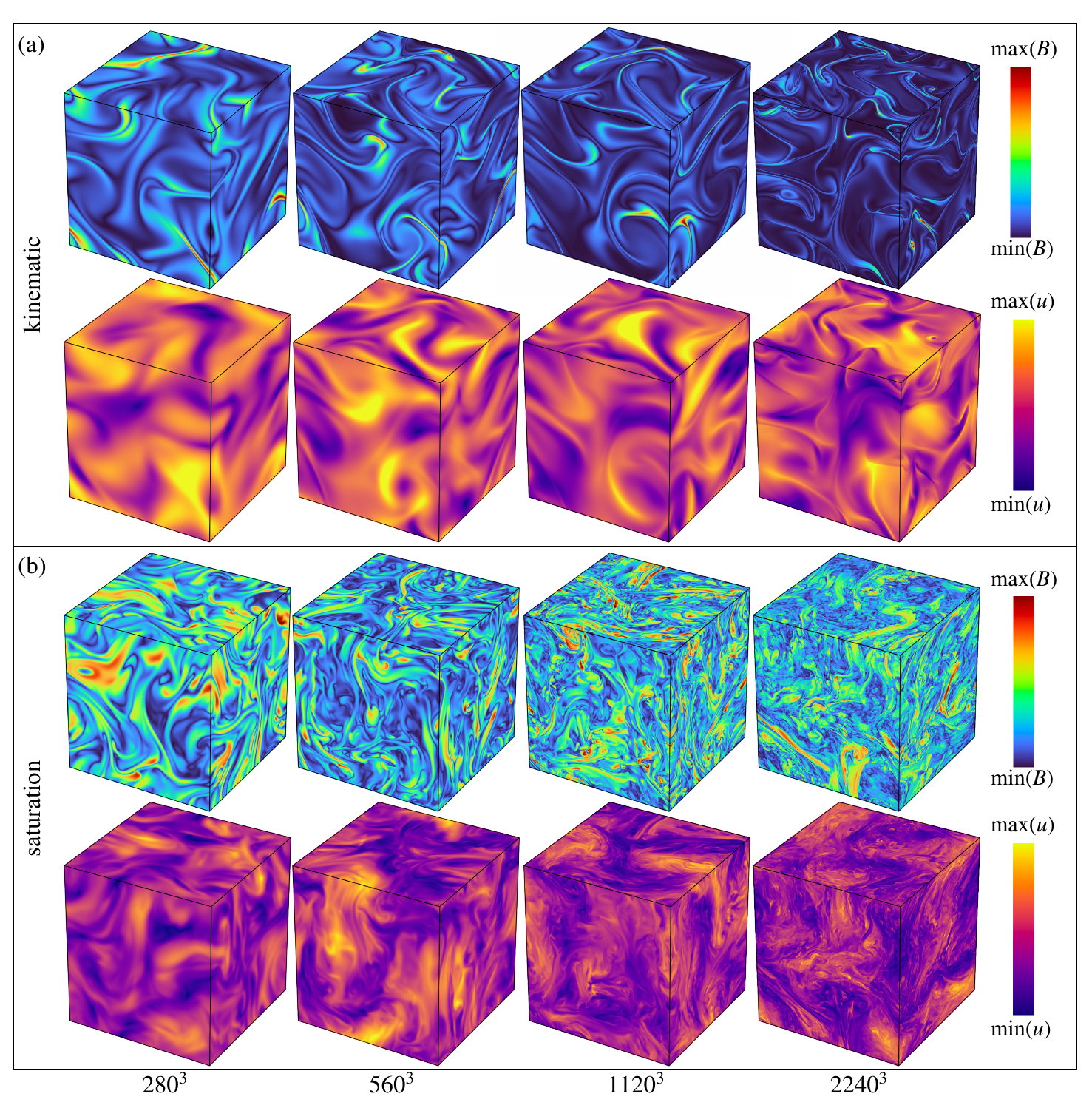}
    \caption{Qualitative comparison of magnetic-field strength $B$ (top row) and speed $u$ (bottom row) during both the kinematic stage (panel~a) and in saturation (panel~b), taken at fixed ${\rm Pm}=10$ from runs a2, b2, c2, and d2. Both quantities are normalized in each snapshot so that the color bars range linearly between their instantaneous minimum and maximum values in the domain. The resolution, and thus ${\rm Rm}$, increases from left to right as indicated.}
    \label{fig:cubes}
\end{figure*}

Figure~\ref{fig:cubes} displays snapshots of the magnetic-field strength $B$ and flow speed $u$ at fixed ${\rm Pm} = 10$ for various ${\rm Rm}$ increasing with resolution. Panel (a) focuses on the end of the kinematic stage, at which point spatially intermittent magnetic folds are readily apparent. As anticipated, there are no clear signs of fold disruption by tearing modes, despite the highly elongated and (particularly at high resolution) thin structures. By contrast, panel (b) shows striking differences in both the magnetic and velocity fields in the saturated state across the sampled range of ${\rm Rm}$. Namely, as ${\rm Rm}$ increases beyond ${\sim}10^4$ (i.e., for resolutions $1120^3$ and above), the magnetic folds are broken up into smaller structures, and the velocity field becomes increasingly filamentary and spatially intermittent (particularly at $2240^3$).

\begin{figure}
    \centering
    \includegraphics[width=\columnwidth]{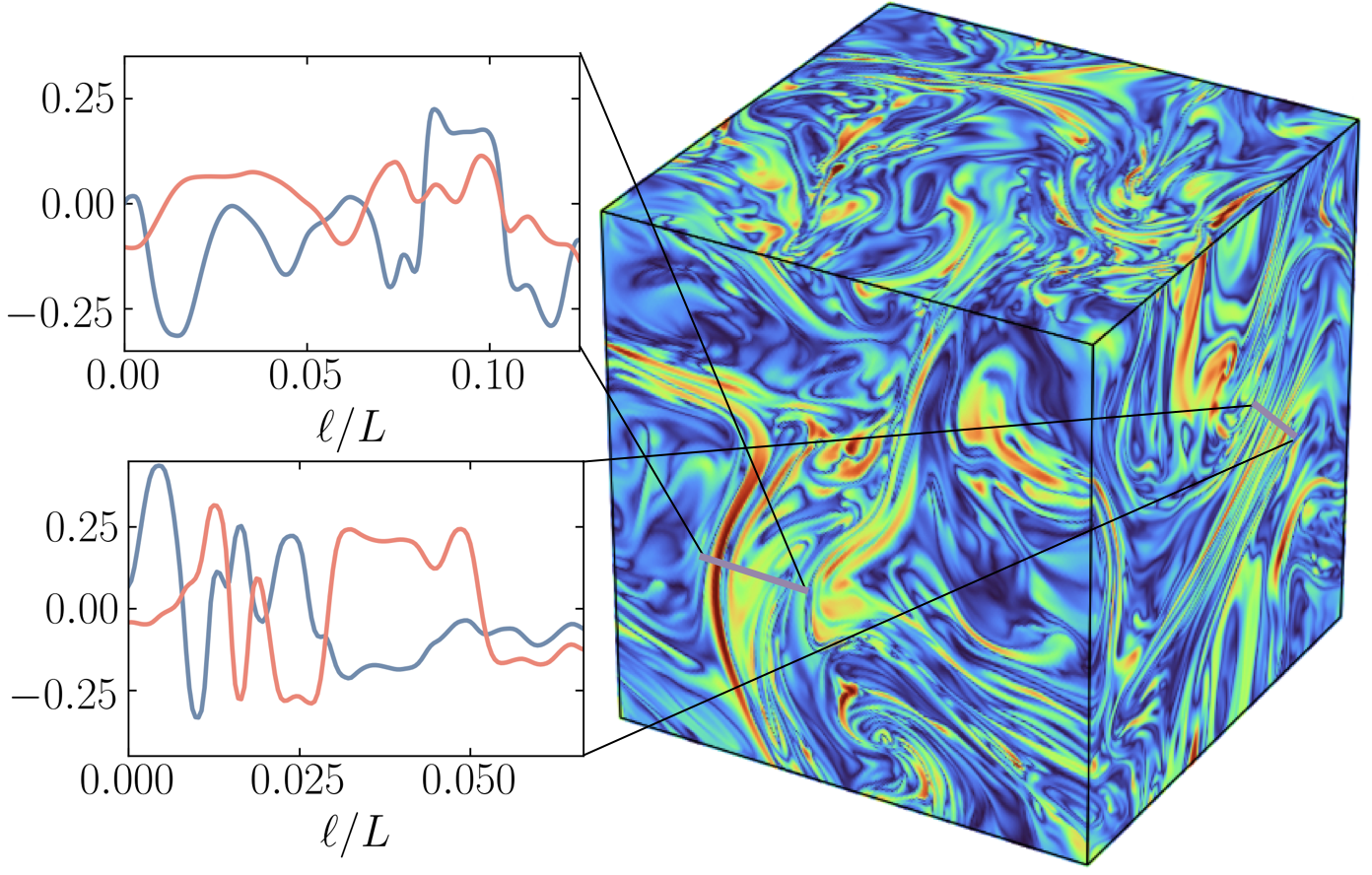}
    \caption{Snapshot of magnetic-field strength in the saturated state of run c7 ($1120^3$, ${\rm Pm} = 500$, ${\rm Re}\approx 20$, ${\rm Rm}\approx 10^4$). Large ${\rm Pm}$ results in elongated laminar current sheets, in contrast to those seen in run c2 ($1120^3$, ${\rm Pm} = 10$, ${\rm Re}\approx 10^3$, ${\rm Rm}\approx 10^4$; see Figure~\ref{fig:cubes}a). Two 1D spatial cuts of the magnetic field are shown on the left side; colors indicate different projections of the magnetic field: in-plane (blue) and out-of-plane (red).}
    \label{fig:Pm500cube}
\end{figure}

\begin{figure*}
    \centering
    \includegraphics[width=\textwidth]{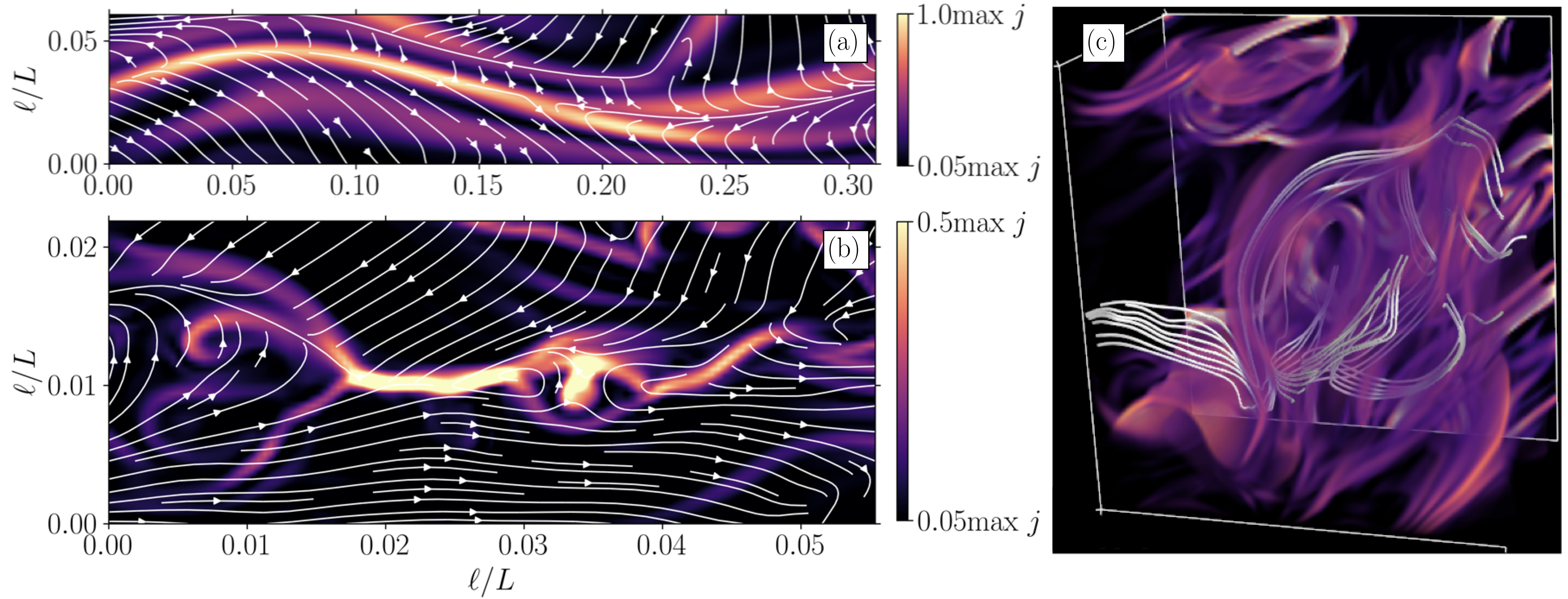}
    \caption{Example of a typical current sheet during saturation for ${\rm Pm} = 10$ at resolution $560^3$ (panel~a) and $2240^3$ (panel~b) (runs b2 and d2, respectively). Color shows the magnitude of the current density; magnetic-field lines are traced by white arrows. Panel~(c) shows a 3D rendering of the current density (color) in a typical plasmoid formed within a disrupted current sheet that is oriented diagonally from bottom left to top right. Magnetic-field lines (white) originate in a small area below and wrap around the central plasmoid; a slice of the current density is also displayed on the back face of the volume.}
    \label{fig:cs}
\end{figure*}

In \S\ref{sec:theory}, we argued that small ${\rm Re}$ (i.e., large ${\rm Pm}$ at fixed ${\rm Rm}$) should suppress the tearing instability, despite the associated increase in the aspect ratio of the folds. Figure~\ref{fig:Pm500cube} provides support for this conjecture, showing a snapshot of the magnetic energy in the saturated state of run c7 ($1120^3$, ${\rm Pm}=500$, ${\rm Re} \approx 20$). Although ${\rm Rm}\approx 10^4$ here is similar to that in the $1120^3$ box with ${\rm Pm}=10$ and ${\rm Re}\approx 10^3$ shown in Figure~\ref{fig:cubes}, the magnetic folds do not appear to be broken up into smaller structures and remain relatively coherent, with strong fields concentrated on small (resistive) scales.

The ${\rm Rm}$ dependence seen qualitatively in Figure~\ref{fig:cubes}(b) is all the more apparent in Figure~\ref{fig:cs}, which provides zoom-ins of individual magnetic structures found in the saturated state of run~b2 ($560^3$) and run~d2 ($2240^3$), both with ${\rm Pm}=10$. The panels show the current density in color and magnetic-field lines in white (with arrows indicating direction). A fairly laminar structure is evident at the lower value of ${\rm Rm}$ (panel~a), with the magnetic field reversing its direction across a smooth current sheet. By contrast, panel~(b) shows a fold obtained at a higher ${\rm Rm}$ that has broken up into smaller current sheets and plasmoid-like structures. A further zoom-in on a similar disrupted fold from run d2 is shown in panel~(c), with the current sheet oriented diagonally from the bottom left to the upper right corner. The magnetic-field lines emanate from a small region near the left side of the central plasmoid and wrap around the central structure, revealing one of the flux ropes.  

In the following sections, these qualitative results and their agreement (or not) with the theoretical arguments of \S\ref{sec:expectations} are made quantitative by examining a variety of diagnostics. When directly comparing to the theory in \S\ref{sec:expectations}, we adopt $n=2$, corresponding to field reversals with sinusoidal (rather than Harris-like) profiles. This choice is supported by the local profiles of the magnetic folds highlighted in Figure~\ref{fig:Pm500cube}, which indicate volume-filling, quasi-sinusoidal variations in the field rather than isolated sheets with tanh profiles. In this case, we predict a noticeable separation between the tearing scale $\lambda_\ast$ and the resistive scale $\ell_\eta$ in the saturated state once ${\rm Re}_{\rm sat}\gtrsim 10^3$ [corresponding to ${\rm Re}^{1/5}_{\rm sat}\gtrsim 4$; see Eq.~\eqref{eqn:amax_sat2} with $n=2$]. Given the values of ${\rm Re}_{\rm sat}$ listed in Table~\ref{tab:runs}, we anticipate clear evidence of tearing-disrupted folds in runs c1, d1, and d2; runs c2 and d3 should be marginal.

\subsection{Energy spectrum and transport}\label{sec:spectrum}

\begin{figure}
    \centering
    \includegraphics[width=\columnwidth]{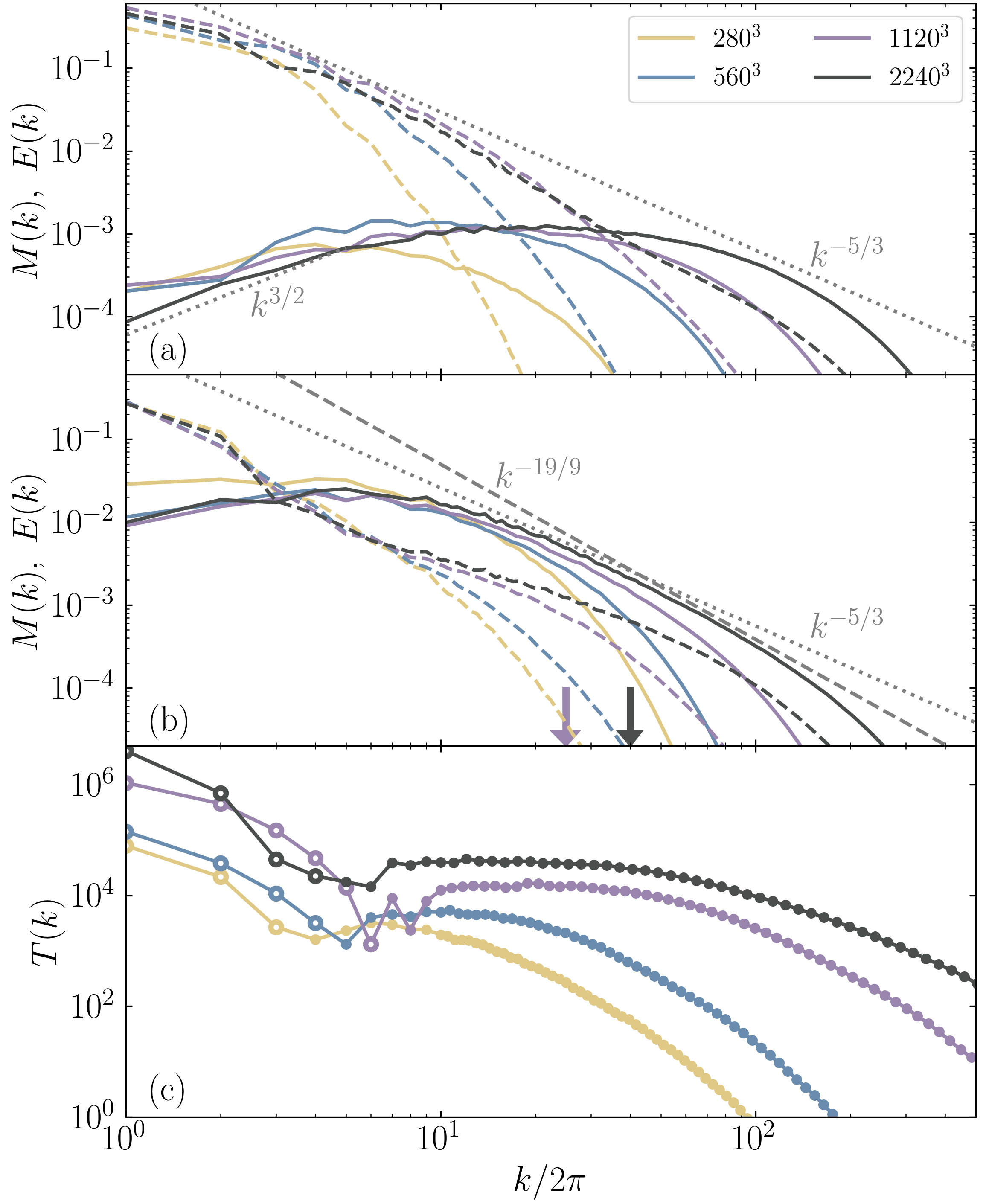}
    \caption{Angle-integrated spectra for ${\rm Pm}=10$ at different resolutions (runs a2, b2, c2, and d2). Color coding of lines is the same in all panels. (a) Kinetic $E(k)$ (dashed) and magnetic $M(k)$ (solid) energy spectra at the end of the kinematic stage, normalized by $\rho_0 u_{\rm rms, sat}^2$. The spectral peak of the magnetic energy depends on the resolution. (b) Kinetic $E(k)$ (dashed) and magnetic $M(k)$ (solid) energy spectra, normalized by $\rho_0 u_{\rm rms, sat}^2$, time-averaged over the saturated state. The arrows indicate the predicted field-reversal scale $\lambda^{-1}_\ast$ [see Eq.~\eqref{eqn:amax_sat}] for ${\rm Pm}=10$ at $1120^3$ (purple) and $2240^3$ (black). The magnetic-energy spectrum acquires a slope steeper than $-5/3$ starting at $k/2\pi \sim \lambda^{-1}_\ast$, which at $2240^3$ is consistent with the spectral envelope of $k^{-19/9}$ expected for a tearing-mediated cascade. The spectral peak of the magnetic energy appears to be independent of ${\rm Rm}$. (c) Transfer function $T(k)$ in saturation, normalized by $\rho_0 u_{\rm rms, sat}^3$, with filled (open) circles corresponding to work done by (against) the Lorentz force.}
    \label{fig:spectrums}
\end{figure}

In this section, we discuss the evolution and parameter dependence of the angle-integrated kinetic and magnetic-energy spectra, which are given respectively by $E(k) \equiv \int\rmd\Omega_k\, k^2 \langle |\bb{u}(\bb{k})|^2\rangle /2$ and $M(k)\equiv \int\rmd\Omega_k \,k^2 \langle |\bb{B}(\bb{k})|^2\rangle /2$~\footnote{The spectra at our highest resolution of $2240^3$ are computed by taking into account every second point in the simulation box, in order to reduce computational cost.}. We also show how energy is pumped into and converted across different scales by the Lorentz force using shell-to-shell transfer functions.

Figure~\ref{fig:spectrums} shows the spectra computed from different resolutions at fixed ${\rm Pm}=10$ (similarly to Figure~\ref{fig:cubes}: runs~a2, b2, c2, and d2) at the end of the kinematic stage (panel~a) and averaged over five snapshots taken during the saturated state (panel~b). During the kinematic stage, the magnetic spectra in all runs (solid lines) follow the expected Kazantsev spectrum ${\propto}k^{3/2}$ at small $k$, with a spectral cutoff that shifts to larger $k$ as ${\rm Rm}$ increases. The latter is qualitatively consistent with arguments made in Section~\ref{sec:expectations} that the kinematic-stage magnetic spectrum should be cut off at a wavenumber ${\propto}{\rm Rm}^{3/4}{\rm Pm}^{1/2}$ (at fixed ${\rm Pm}$ in this case). As the magnetic energy builds up exponentially, the Lorentz force eventually becomes large enough to back-react on the flow at the viscous scale, steepening the kinetic-energy spectrum (dashed lines) in comparison with its being approximately Kolmogorov's $k^{-5/3}$ at large scales.

During the nonlinear stage, when the magnetic energy exhibits secular growth, the spectral peak of $M(k)$ shifts towards smaller wavenumbers (larger scales), as expected theoretically. However, this peak follows neither $\ell^{-1}_\eta \propto {\rm Rm}^{1/2}$ nor $\lambda_\ast^{-1} \propto {\rm Rm}^{3/10} {\rm Pm}^{1/5}$ in the saturated state (panel~b), but rather appears to be independent of (or, at most, very weakly dependent on) ${\rm Rm}$ at the higher resolutions. This point is revisited in \S\ref{sec:scale}, where we compute the integral scale of the magnetic field and show quantitatively that it becomes approximately independent of ${\rm Rm}$ in the saturated state for ${\rm Rm}\gtrsim 10^4$, ${\rm Pm}\lesssim 50$.

On scales smaller than that on which the magnetic spectrum peaks, $M(k)$ steepens gradually and, at the highest resolution, appears to acquire a power law that is steeper than Kolmogorov but consistent with the $k^{-19/9}$ envelope predicted at the end of \S\ref{sec:theory} for a cascade controlled by tearing (the scale separation between $\lambda_\ast$ and $\ell_{\eta}$ is not large enough to determine definitively the exact spectral index). 
%On scales smaller than that on which the magnetic spectrum peaks, $M(k)$ steepens gradually and, at the highest resolution, ultimately acquires a power law that is steeper than Kolmogorov but consistent with the $k^{-19/9}$ envelope predicted at the end of \S\ref{sec:theory} for a cascade controlled by tearing.
It is notable that the consistency with $-19/9$ begins around $k/2\pi \sim 40$ at $2240^3$, which, perhaps not coincidentally, matches the predicted value of $\lambda_\ast^{-1}$ given ${\rm Pm}=10$ and the measured ${\rm Rm}_{\rm sat}$ for this run [indicated by the black arrow; see Eq.~\eqref{eqn:tearingscale}]. At lower resolutions, ${\rm Re}^{1/5}_{\rm sat}$ is not sufficiently large for tearing to act faster than resistive decay, with the situation at $1120^3$, ${\rm Pm}=10$ being marginal at best (purple arrow).

These changes in the magnetic- and kinetic- energy spectra reflect not only the disruption of magnetic folds by the tearing instability, but also the modified interplay between the flow and the Lorentz force. This interplay may be studied using the local shell-to-shell transfer function, $T(k) = \int\rmd\Omega_k \, k^2 \bb{u}(\bb{k}) [\bb{J}\btimes\bb{B}](\bb{k})^\ast$~\cite{Brandenburg2019,Grete2021}. This function describes the amount of work done by ($T>0$) or against ($T<0$) the Lorentz force at a single scale $k$. Following Ref.~\cite{Brandenburg2019}, we associate $T(k)<0$ with ``forward dynamo action'', corresponding to growth of the magnetic energy at the expense of the kinetic energy; and $T(k)>0$ with ``reversed dynamo action'', corresponding to a transfer from magnetic energy to kinetic energy. Figure~\ref{fig:spectrums}(c) shows $T(k)$ for the same runs as in panels (a) and (b). At the largest scales, where $E(k)>M(k)$, energy injection is working against the Lorentz force, corresponding to forward dynamo action (denoted by open circles). Work done by the Lorentz force increases gradually towards smaller scales until $E(k) \sim M(k)$, beyond which $T(k)$ becomes positive (filled circles). For ${\rm Rm} \gtrsim 10^4$ (resolutions of $1120^3$ and $2240^3$), there is a wavenumber range in which $T(k)$ is roughly constant, corresponding to a constant magnetic-to-kinetic energy flux. We think that it is in this range that the coherent magnetic folds are driving flows by exerting a Lorentz force. That process becomes more difficult if these coherent folds break up into plasmoid-like flux ropes~\cite{blackman96}; and so it may be no coincidence that the transfer function stops being constant in $k$ right at the wavenumber $k=2\pi/\lambda_\ast$ where tearing is predicted to onset, and the measured $M(k)$ steepens to a spectrum consistent with $k^{-19/9}$.

\begin{figure}
    \centering
    \includegraphics[width=\columnwidth]{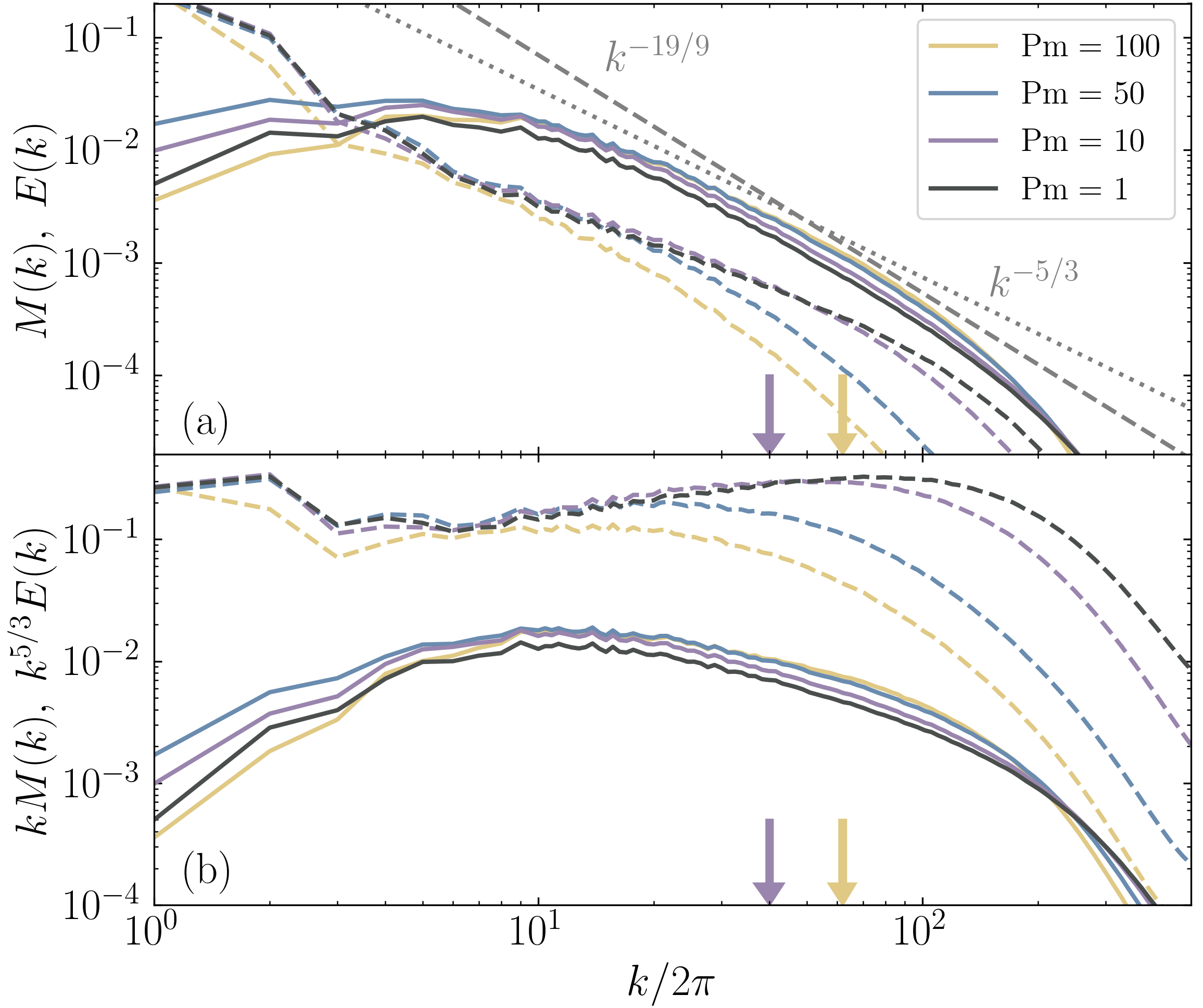}
    \caption{(a) Angle-integrated kinetic $E(k)$ (dashed) and magnetic $M(k)$ (solid) spectra, time-averaged over the saturated state of runs d1, d2, d3, and d4 ($2240^3$ at different ${\rm Pm}$), normalized by $\rho_0 u_{\rm rms, sat}^2$. The magnetic-energy spectrum steepens to be consistent with the predicted $-19/9$ envelope for a tearing-mediated cascade at a wavenumber that increases slightly with ${\rm Pm}$; cf.~Eq.~\eqref{eqn:amax_sat}, which predicts $\lambda^{-1}_\ast \propto {\rm Pm}^{1/5}$ at fixed ${\rm Rm}$ when $n=2$. The arrows indicate the predicted $\lambda^{-1}_\ast$ for ${\rm Pm}=10$ (purple) and $100$ (yellow). (b) Kinetic spectra, compensated by $k^{5/3}$, and magnetic spectra, compensated by $k$, to illustrate the argument made at the end of \S\ref{sec:scale}; here $kM(k)$ is multiplied by an arbitrary factor of 0.1 to separate it visually from the other curves.}
    \label{fig:2240spec}
\end{figure}

To close this subsection, we provide in Figure~\ref{fig:2240spec} the kinetic- and magnetic-energy spectra at $2240^3$ (our largest ${\rm Rm}$) for ${\rm Pm}=\{1,10,50,100\}$, time-averaged over the saturated state. A close examination reveals a slight increase with ${\rm Pm}$ of the wavenumber at which the magnetic spectrum steepens to be consistent with the predicted $-19/9$ envelope for a tearing-mediated cascade. In particular, at ${\rm Pm}=100$, $M(k)$ remains no steeper than $k^{-5/3}$ until $k/2\pi \gtrsim 60$. Substituting the values of~${\rm Re}_{\rm sat}$ and ${\rm Rm}_{\rm sat}$ listed in Table~\ref{tab:runs} for run d4 into Eq.~\eqref{eqn:amax_sat} with $n=2$ yields $\lambda^{-1}_\ast \approx 62$ (yellow arrow), consistent with this observation.

\subsection{Characteristic scales of the magnetic field}\label{sec:scale}

\begin{figure}
    \centering
    \includegraphics[width=\columnwidth]{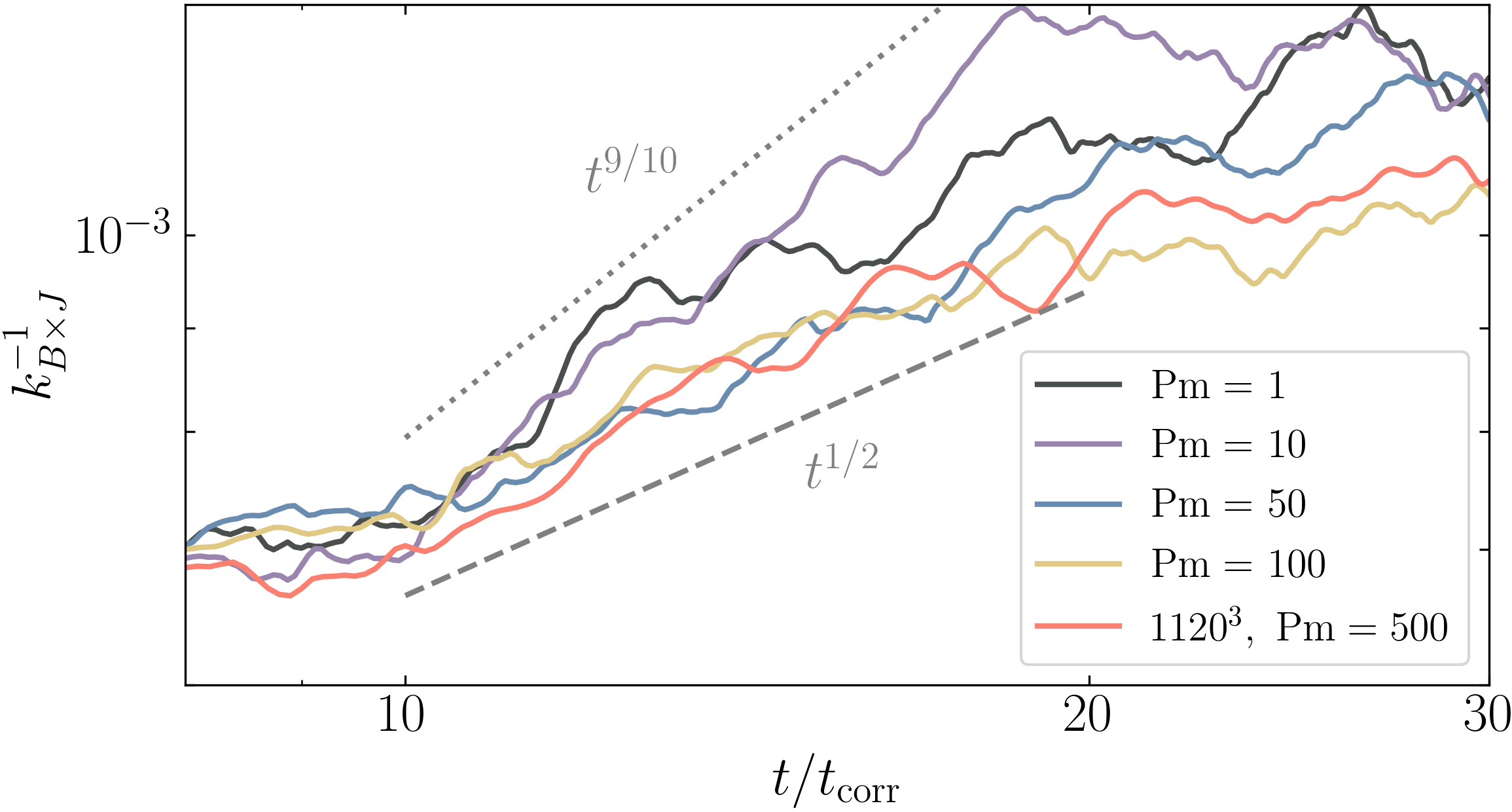}
    \caption{Time evolution of the field-reversal scale $\kbcj^{-1}$ during the nonlinear stage at resolution $2240^3$ for different ${\rm Pm}$ (runs d1, d2, d3, d4) and at resolution $1120^3$ with ${\rm Pm =500}$ (run c7; here $\kbcj$ is multiplied by an arbitrary factor of $1.4$ to bring it nearer visually to the other curves).}
    \label{fig:frs_evolution}
\end{figure}

\begin{figure}
    \centering
    \includegraphics[width=\columnwidth]{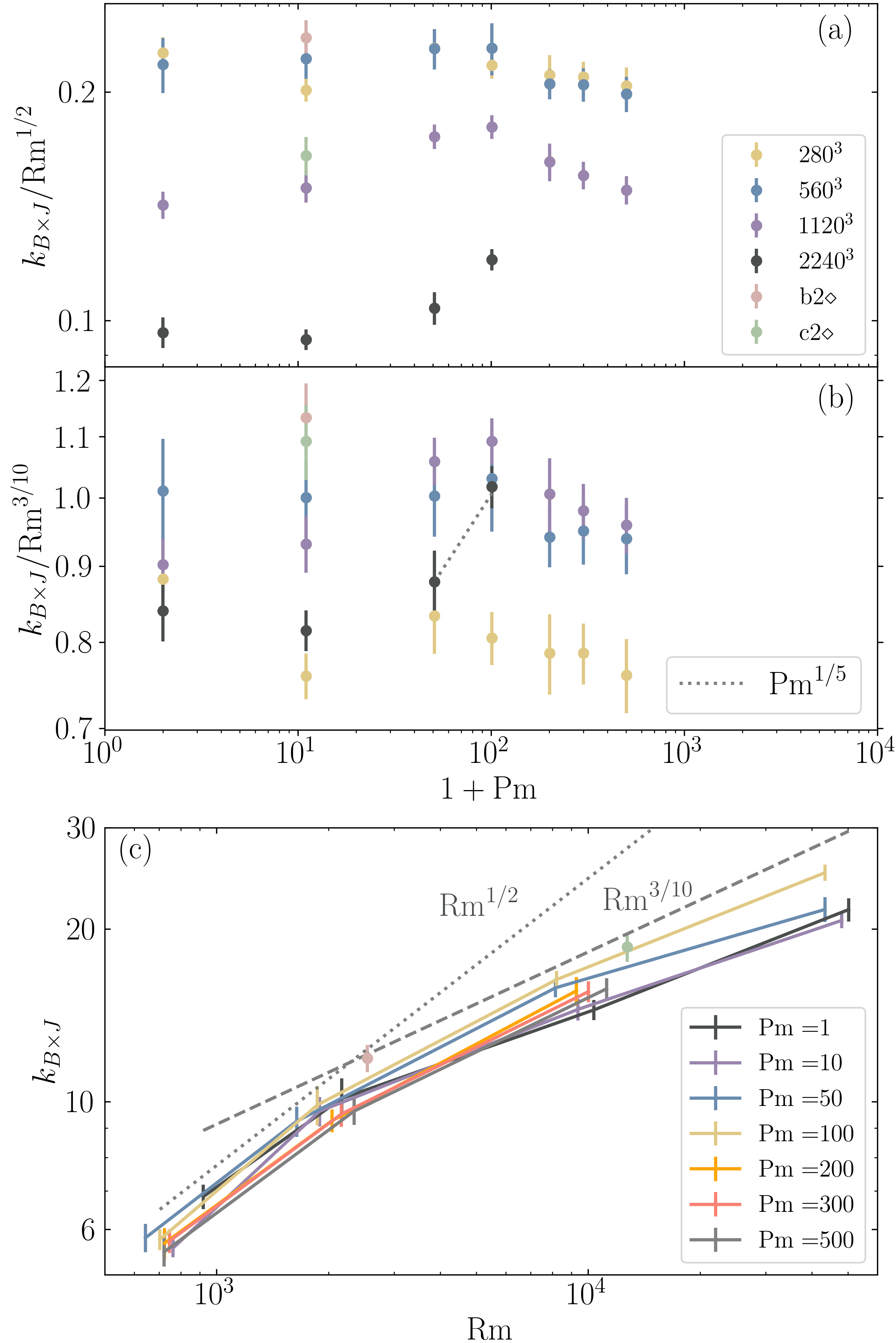}
    \caption{Dependence of the field-reversal wavenumber $\kbcj$ on plasma parameters in the saturated state. (a,b) $\kbcj$ vs.~${\rm Pm}$ at different ${\rm Rm}$; the color coding is the same for these two panels. In panel (a), $\kbcj$ is divided by ${\rm Rm}^{1/2}$, the predicted ${\rm Rm}$ dependence of the inverse resistive scale $\ell^{-1}_\eta$; in panel (b), $\kbcj$ is divided by ${\rm Rm}^{3/10}$, the predicted ${\rm Rm}$ dependence of the inverse tearing scale $\lambda^{-1}_\ast$ [see Eq.~\eqref{eqn:tearingscale}]. (c) $\kbcj$ vs.~${\rm Rm}$ at different $\rm Pm$. Low-${\rm Rm}$ runs show an ${\rm Rm}$ dependence consistent with the resistive scale, $\kbcj\sim 2\pi/\ell_\eta$; high-${\rm Rm}$ runs show an ${\rm Rm}$ and ${\rm Pm}$ dependence consistent with the predicted tearing scale, $\kbcj\sim 2\pi/\lambda_\ast$.}
    \label{fig:frs}
\end{figure}

The geometry of the magnetic field can be further quantified using the following characteristic wavenumbers~\cite{Scheko04sim}:
\begin{align}
    k_{\parallel} &\equiv \left( \frac{\langle |\bb{B}\bcdot\grad\bb{B}|^2 \rangle}{\langle B^4 \rangle} \right)^{1/2}, \\*
    \kbdj &\equiv \left( \frac{\langle|\bb{B}\bcdot\bb{J}|^2\rangle}{\langle B^4\rangle} \right)^{1/2}, \\*
    \kbcj &\equiv \left( \frac{\langle |\bb{B}\btimes\bb{J} |^2 \rangle}{\langle B^4 \rangle} \right)^{1/2},
    \label{eqn:characteristic_scales}
\end{align}
where $\langle\, \dots\rangle$ denotes a statistical (box) average. These wavenumbers measure the characteristic variation of the magnetic field along itself (``$\parallel$'') and across itself, with the latter two perpendicular directions oriented using the local direction of the current density $\bb{J}$. Following Ref.~\cite{Scheko04sim}, we associate with these scales the characteristic length, width, and thickness (in that order) of the magnetic folds. The final wavenumber, $\kbcj$, is particularly important, as it quantifies the characteristic reversal scale of the magnetic field. During the kinematic stage, we expect $\kbcj\ell_\eta\sim 1$, whereas during the nonlinear stage and saturated state, $\kbcj\lambda_\ast\sim 1$. Thus, if tearing is important, then we predict $\kbcj\propto t^{9/10}$ during the nonlinear stage and $\kbcj \propto {\rm Rm}^{3/10}{\rm Pm}^{1/5}$ in saturation; if tearing is not important, then these scalings become $t^{1/2}$ and ${\rm Rm}^{1/2}$, respectively.

In Figure~\ref{fig:frs_evolution}, we examine the time evolution of $\kbcj$ during the nonlinear stage using results at $2240^3$ for ${\rm Pm}=\{1,10,50,100\}$ and at $1120^3$ for ${\rm Pm}=500$. At large values of ${\rm Pm}$, we obtain evolution consistent with a field-reversal scale that is proportional to the resistive scale, $\kbcj\propto t^{1/2}$. However, at ${\rm Pm}=10$, the evolution is much closer to the tearing scale, with $\kbcj\propto t^{9/10}$. Again, this particular value of ${\rm Pm}$ at $2240^3$ is notable in that it is large enough to facilitate the production of large-aspect-ratio current sheets but not so large as to interfere with the disruption of such sheets by tearing.

The resistive scale predicted for the saturated state of the dynamo was given in~\S\ref{sec:folds} as $\ell_{\eta} \sim L\,{\rm Rm}^{-1/2}$, independent of ${\rm Re}$. Therefore, if the reversal scale were set by $\ell_\eta$, the ratio $\kbcj/{\rm Rm}^{1/2}$ should stay roughly constant as ${\rm Re}$ and ${\rm Rm}$ are varied across our parameter scan. Indeed, Figure~\ref{fig:frs}(a) demonstrates little variation in $\kbcj/{\rm Rm}^{1/2}$ across all ${\rm Pm}$ when ${\rm Rm}$ takes on relatively small values (at resolutions $280^3$ and $560^3$, corresponding to runs a1--a7 and b1--b7, respectively). This scaling is consistent with the results of other published ${\rm Pm}\gtrsim 1$ dynamo simulations that had relatively low resolutions (see, e.g., figure 16 of Ref.~\cite{Scheko04sim}). However, at higher resolutions ($1120^3$ and especially $2240^3$), the values of~$\kbcj$ depart from this scaling once ${\rm Pm}\gtrsim 10$: we now see $\kbcj/{\rm Rm}^{1/2}$ increasing with ${\rm Pm}$. Such a dependence on viscosity is consistent with the idea that, at sufficiently high ${\rm Rm}$ and intermediate values of ${\rm Pm}$, magnetic folds should have their reversals limited by the tearing scale $\lambda_\ast\propto {\rm Rm}^{-3/10}{\rm Pm}^{-1/5}$ rather than by the resistive scale~$\eta_\eta$. Indeed, if we instead normalize $\kbcj$ using ${\rm Rm}^{3/10}$ as in Figure~\ref{fig:frs}(b), we see behavior consistent with ${\rm Pm}^{1/5}$ at our largest resolutions. Unfortunately, at present, the steep numerical cost does not allow us to verify this dependence at yet higher resolutions and larger values of ${\rm Pm}$, and so we view these trends as consistent with our theory rather than as confirmatory in a definitive way.

In Figure~\ref{fig:frs}(c), these results are re-organized: $\kbcj$ is plotted versus ${\rm Rm}$, with different ${\rm Pm}$ indicated by the different colors. As in the previous two panels, there is a general trend away from the ${\rm Rm}^{1/2}$ scaling at low ${\rm Rm}$ and towards one consistent with ${\rm Rm}^{3/10}$ at high ${\rm Rm}$. We take this as our clearest quantitative evidence for a change in the characteristic field-reversal scale due to tearing at high ${\rm Rm}$. Note that $\kbcj \gtrsim 20$ at our highest resolution, implying Lundquist numbers ${\sim}B_{\rm rms, sat}/\kbcj \eta \gtrsim 10^4$.

\begin{figure}
    \centering
    \includegraphics[width=\columnwidth]{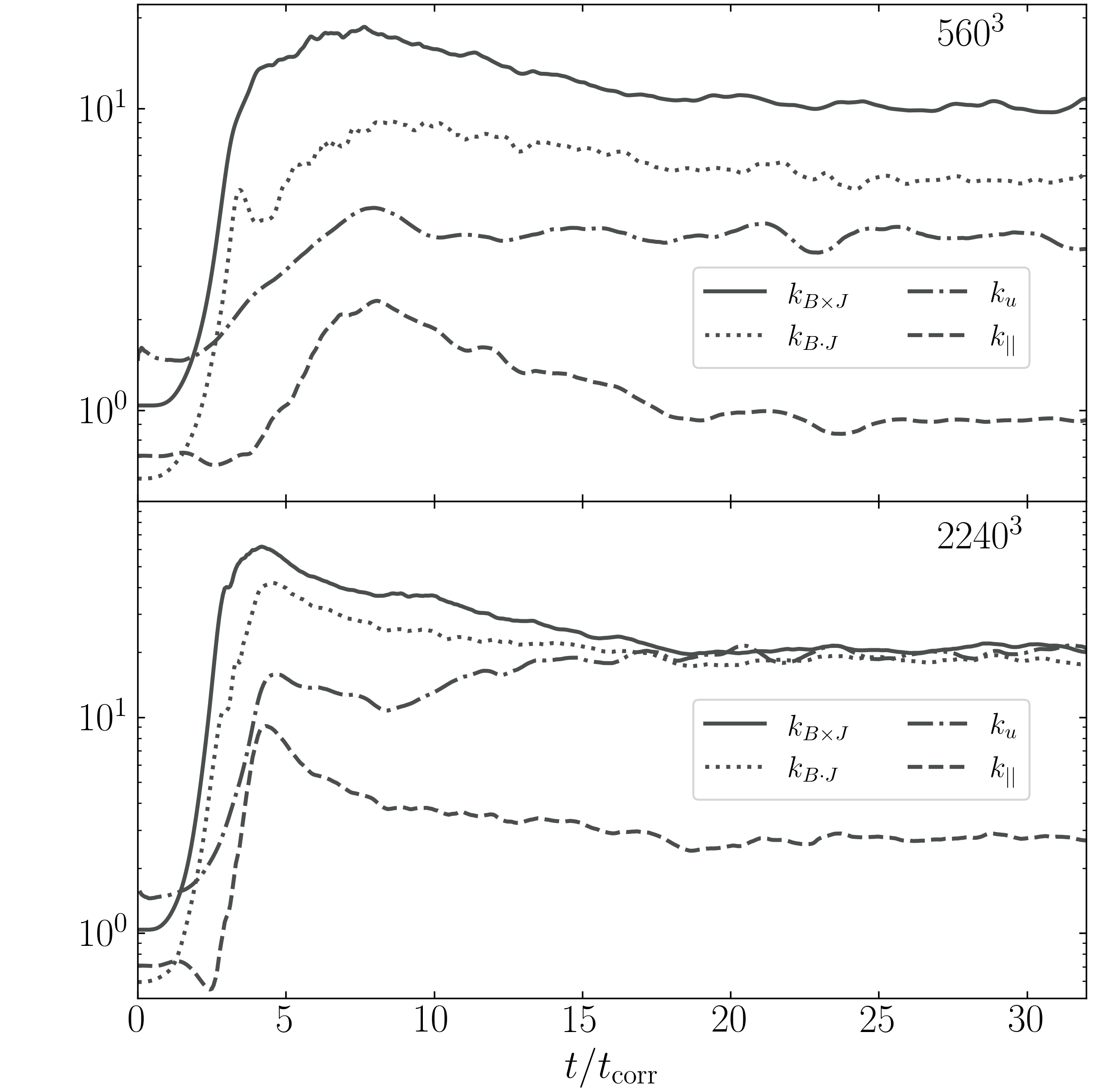}
    \caption{Time evolution of the characteristic wavenumbers $(k_\parallel,\kbdj,\kbcj)$ describing the magnetic field [see Eq.~\eqref{eqn:characteristic_scales}] and $k_u$ describing the small-scale structure of the velocity field [see Eq.~\eqref{eqn:klambda}] at ${\rm Pm}=10$ and resolutions $560^3$ and $2240^3$.
    }
    \label{fig:allk}
\end{figure}

The time evolution of the two other characteristic wavenumbers of the magnetic field, $k_\parallel$ and $\kbdj$, is shown alongside that of $\kbcj$ in Figure~\ref{fig:allk} at ${\rm Pm}=10$ and resolutions $560^3$ and $2240^3$. At low resolution, these wavenumbers have a clear ordering from the kinematic phase all the way through to saturation: $k_\parallel \ll \kbdj < \kbcj$, with the latter two becoming closer to one another than the former two in saturation ({\em viz.}, $k_\parallel : \kbdj : \kbcj \approx 1 : 7 : 12$; the exact numbers depend on ${\rm Rm}$). Such a hierarchy suggests folded magnetic fields organized into flux ribbons. In the saturated state at high resolution, this arrangement is modified quite dramatically: the two perpendicular scales become comparable to one another, $k_\parallel \ll \kbdj \simeq \kbcj$. This is instead indicative of folded magnetic fields organized into flux ropes, a natural outcome of plasmoid formation during the tearing disruption of current sheets. Interestingly, the rms wavenumber of the velocity field in this high-resolution run (the ``Taylor microscale''),
\begin{equation}\label{eqn:klambda}
    k_u \equiv \left( \frac{\langle |\grad\bb{u}|^2 \rangle}{\langle u^2 \rangle} \right)^{1/2} ,
\end{equation}
increases during the nonlinear stage to become comparable to $\kbdj$ and $\kbcj$, indicating flows on the flux-rope (field-reversal) scale. That this does {\em not} occur at low resolution is further support for the tearing disruption of folds at high resolution.

\begin{figure}
    \includegraphics[width=\columnwidth]{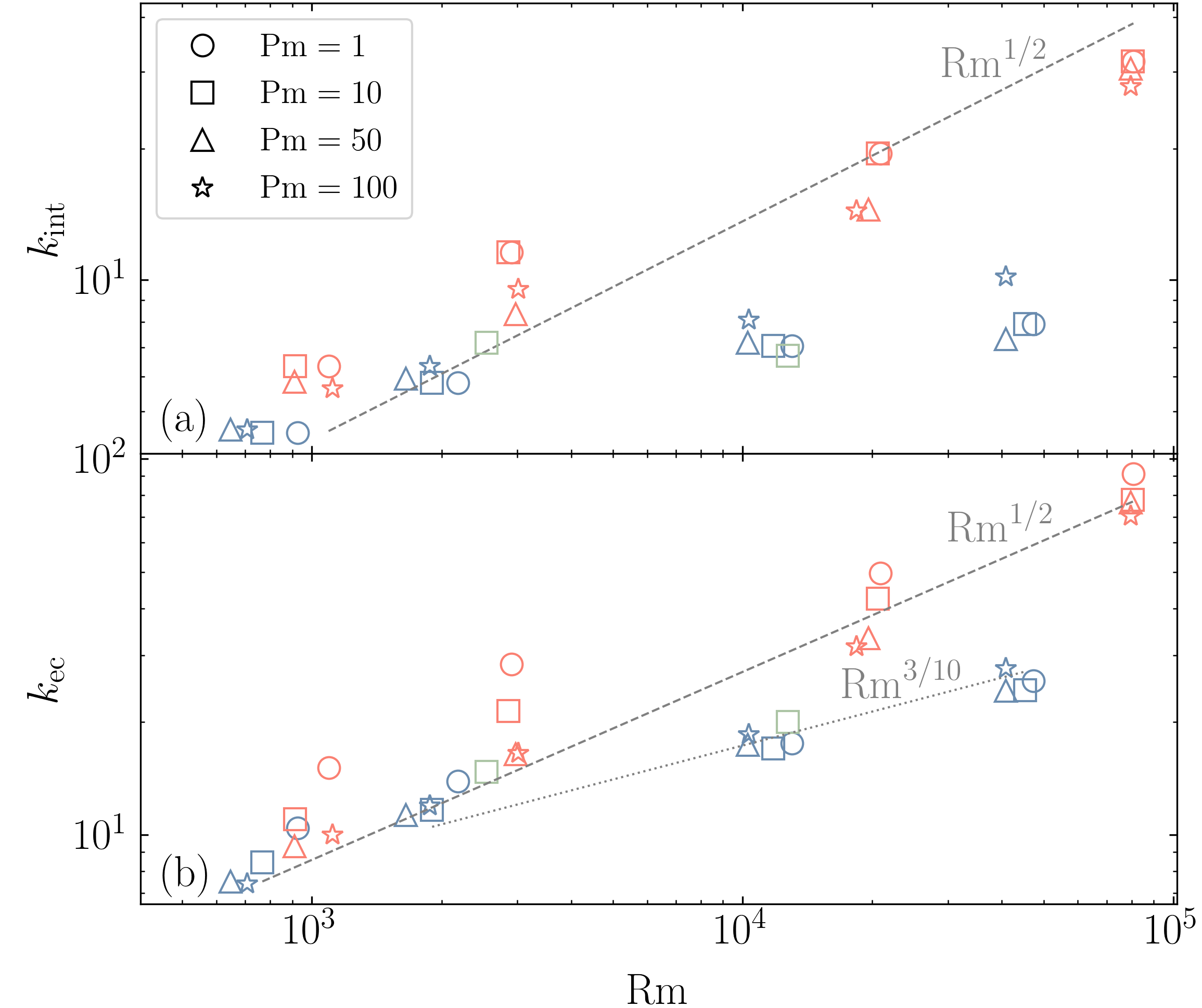}
    \caption{Dependence of (a) the integral scale $k_{\rm int}$ [Eq.~\eqref{eqn:kint}] and (b) the energy-containing scale $k_{\rm ec}$ [Eq.~\eqref{eqn:kec}] on ${\rm Rm}$ at the end of the kinematic stage (red markers) and in saturation (blue markers). Different markers correspond to different ${\rm Pm}$, as indicated; the green squares refer to the saturated state of runs b2$\diamond$ and c2$\diamond$. At the end of the kinematic stage, both $k_{\rm int}$ and $k_{\rm ec}$ vary approximately as ${\rm Rm}^{1/2}$, consistent with $k_{\rm int}\sim k_{\rm ec} \sim \ell^{-1}_\eta$. In saturation, $k_{\rm int}$ appears to approach a constant value at large ${\rm Rm}$, at least for ${\rm Pm}\le 50$, while $k_{\rm ec}$ becomes consistent with the tearing scale, ${\propto}{\rm Rm}^{3/10}$.}
    \label{fig:ecs}
\end{figure}

Lastly, we follow up on the discussion of the magnetic spectrum in \S\ref{sec:spectrum}, and analyze quantitatively its (conventionally defined) ``integral scale''
\begin{equation}\label{eqn:kint}
    k_{\rm int} \equiv \int\rmd k \, M(k) \bigg/ \int\rmd k \, k^{-1} M(k) ,
\end{equation}
which is a reasonable proxy for the scale of the spectral peak. For $M(k)$ scaling as $k^\alpha$ with $\alpha\not\in[-1,0]$, $k_{\rm int}$ is also the energy-containing scale, i.e., the peak of $kM(k)$. As was discussed in Sec.~\ref{sec:folds}, the magnetic energy is concentrated near the resistive scale $\ell_\eta \propto {\rm Rm}^{-1/2}$ throughout the kinematic stage and subsequently moves to larger scales during the secular stage of evolution.

Figure~\ref{fig:ecs}(a) shows the dependence of $k_{\rm int}$ on ${\rm Rm}$ obtained from all our simulations during the kinematic stage (red symbols) and in the saturated state (blue symbols); the ${\rm Rm}^{1/2}$ scaling expected to hold during the kinematic stage is overlaid for comparison. At all resolutions and for all values of ${\rm Pm}$, the ${\rm Rm}^{1/2}$ scaling appears to be well satisfied at the end of the kinematic stage. In saturation, the values of $k_{\rm int}$ calculated at resolutions $280^3$ and $560^3$ (${\rm Rm}\lesssim 2\times 10^3$) also satisfy this scaling. At resolutions $1120^3$ and $2240^3$ (${\rm Rm}\gtrsim 10^4$), however, this trend breaks and $k_{\rm int}$ measured in saturation becomes nearly independent of ${\rm Rm}$ as ${\rm Rm}$ increases. Confirming that $k_{\rm int}$ becomes truly independent of ${\rm Rm}$ awaits higher-resolution simulations, but our results so far strongly suggest that this is the case. 

This might appear to match the expectations of a number of previous authors~\cite{bs51,subramanian99,haugen03,haugen04,beresnyak12,XuLazarian_2016}, but it is important to understand the distinction between the scale of the spectral peak and the energy-containing scale of the magnetic field, which is the peak of $kM(k)$. The situation with the latter is (even) murkier than with $k_{\rm int}$. Analogously to Eq.~\eqref{eqn:kint}, a proxy for the energy-containing scale can be defined as
\begin{equation}\label{eqn:kec}
    k_{\rm ec} \equiv \int\rmd k \, kM(k) \bigg/ \int\rmd k \, M(k) .
\end{equation}
The dependence of $k_{\rm ec}$ on ${\rm Rm}$ and ${\rm Pm}$ for all our runs is shown in Figure~\ref{fig:ecs}(b). Again, at lower ${\rm Rm}$, $k_{\rm ec}\propto{\rm Rm}^{1/2}$ works well, whereas at higher ${\rm Rm}$, the scaling with ${\rm Rm}$ becomes weaker but does not entirely flatten---indeed, it may be consistent with $k_{\rm ec}\propto \lambda^{-1}_\ast \propto {\rm Rm}^{3/10}{\rm Pm}^{1/5}$. If this is indeed true, it would suggest that, at $k\lambda_\ast\lesssim 1$, $M(k)\propto k^{-1}$---equivalently, the mean squares of magnetic increments across any distance $\lambda>\lambda_\ast$ would all have the same value, a possibility mooted by Ref.~\cite{Yousef2007} for a saturated folded field. While the scale separation between $\lambda_\ast$ and the system size $L$ in our simulations is not large enough to tease out any such scaling definitively, plotting $kM(k)$ in Figure~\ref{fig:2240spec}(b) confirms that a $k^{-1}$ scaling of the magnetic spectrum cannot, at these resolutions, be ruled out.

\subsection{Viscous and resistive dissipation}

\begin{figure}
    \centering
    \includegraphics[width=\columnwidth]{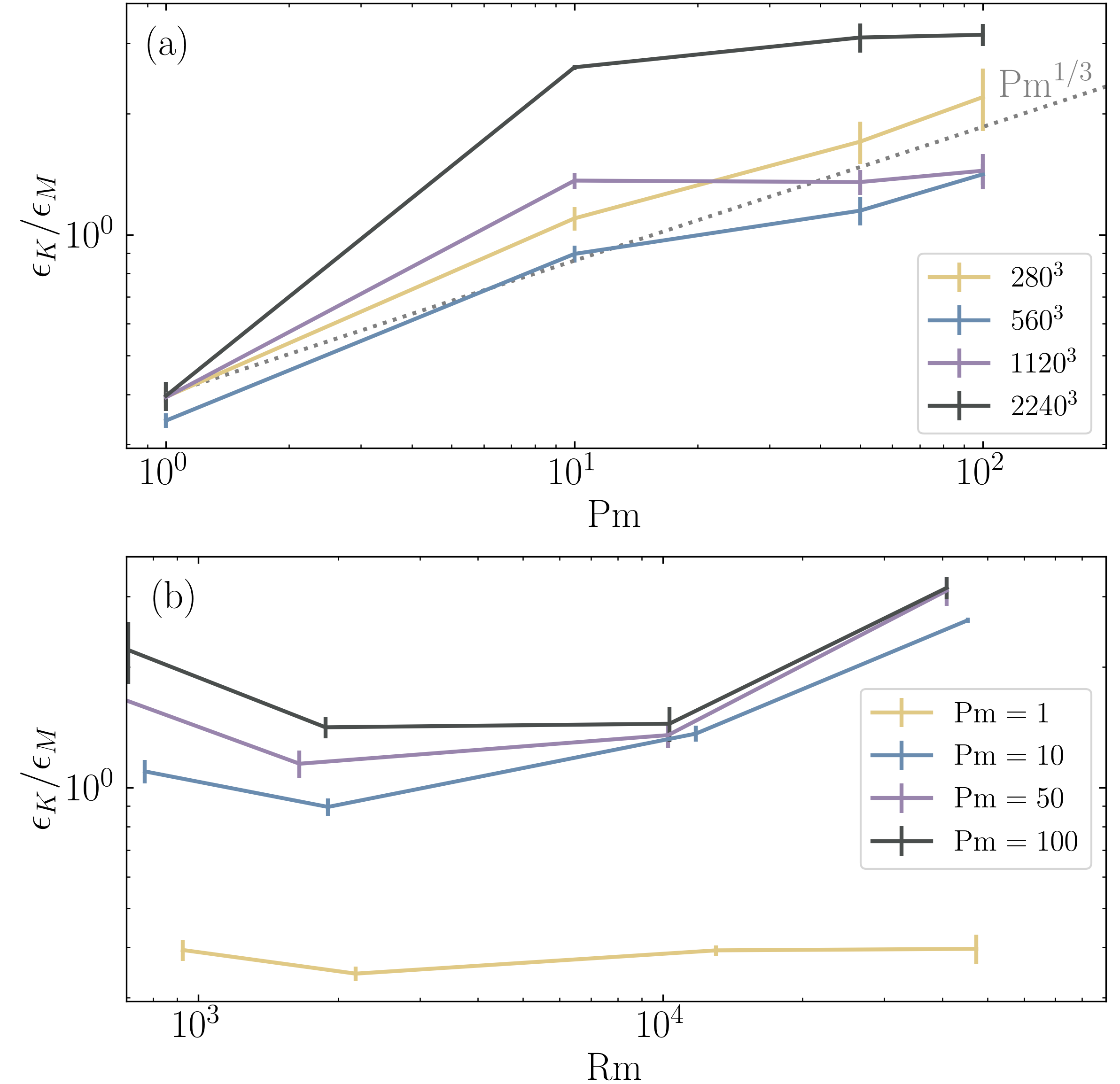}
    \caption{Ratio of turbulent energy dissipated viscously versus resistively in the saturated state (a) versus ${\rm Pm}$ at different resolutions (and therefore different ${\rm Rm}$) and (b) versus ${\rm Rm}$ at different ${\rm Pm}$. Error bars reflect the standard deviation in time.}
    \label{fig:heating}
\end{figure}

To close this section, we compute the box-averaged dissipation rates of the kinetic and magnetic energies,
\begin{equation}
    \epsilon_K \equiv \nu \langle|\grad\bb{u}|^2 \rangle \quad{\rm and}\quad \epsilon_M \equiv \eta \langle|\bb{J}|^2\rangle ,
\end{equation}
respectively, and average them over the saturated state. Figure~\ref{fig:heating} presents the ratio of energy dissipated viscously versus resistively, $\epsilon_K/\epsilon_M$, (a) as a function of ${\rm Pm}$ for different resolutions (and therefore different ${\rm Rm}$) and (b) as a function of ${\rm Rm}$ at fixed ${\rm Pm}$. At low and moderate resolutions and for ${\rm Pm}\le 10$, $\epsilon_K/\epsilon_M$ is approximately proportional to ${\rm Pm}^{1/3}$ (dotted line), an empirical scaling reported previously in Refs.~\cite{Brandenburg2014,McKay2019}. For ${\rm Pm} > 10$, this ratio becomes roughly independent of ${\rm Pm}$ at fixed ${\rm Rm}$ for our two highest resolutions. The dissipation ratio at fixed ${\rm Pm}$ also increases from ${\rm Rm} \approx 10^4$ to ${\rm Rm} \approx 4\times 10^4$, after being relatively flat for smaller values of ${\rm Rm}$. We attribute this increase in viscous dissipation at large ${\rm Rm}$ to the tearing-initiated breakup of intense current sheets leading to small-scale reconnection outflows and intra-island circulations, features that will be explored in a separate publication alongside a detailed analysis of the flow and field statistics in saturation.

\section{Summary and outlook}\label{sec:summary}

Using analytical arguments and high-resolution, viscoresistive MHD simulations at ${\rm Pm} \ge 1$, we have demonstrated that the elongated magnetic folds naturally produced by the turbulent dynamo become unstable to tearing during the nonlinear stage and in the saturated state once ${\rm Re}^{1/5}\gg 1$. During the kinematic stage, the tearing instability of these folds cannot grow fast enough to overcome their resistive decay, a prediction confirmed by our numerical simulations. As a result, current sheets are visually featureless and the magnetic energy resides at the smallest available scale (resistive) until the magnetic field becomes strong enough to back-react on the flow. Thereafter, the current sheets begin to break up into plasmoid-like flux ropes on increasingly larger scales. The characteristic field-reversal scale in the saturated state changes from the standard resistive ${\rm Rm}^{1/2}$ scaling towards one consistent with ${\rm Rm}^{3/10}{\rm Pm}^{1/5}$, as predicted from a balance between the characteristic linear tearing timescale of sinusoidal folds and the nonlinear turnover time at the outer scale.

Other quantities calculated during the saturated state at large ${\rm Rm}$ are also consistent with our expectations for a tearing-limited dynamo. The magnetic spectrum appears to steepen below the anticipated maximal tearing scale to take on a spectral index consistent with $-19/9$, the spectral envelope theoretically predicted for tearing-mediated Alfv\'{e}nic turbulence (albeit over an uncomfortably short wavenumber range due to limited resolution). And the characteristic reversal scale and width of the magnetic folds become comparable to one another, a feature not seen at lower values of ${\rm Rm}$ and consistent with flux ropes being produced by the tearing-induced disruption of thin current sheets. That the Taylor microscale of the velocity field decreases during the nonlinear stage of the dynamo to become comparable to these fold scales supports this scenario. The implied sub-viscous injection of kinetic energy by the Lorentz force that occurs at these large values of ${\rm Rm}$ is found to result in a pronounced excess of viscous dissipation over resistive dissipation for ${\rm Pm}\ge 10$.

Finally, we have found that the spectral peak of the magnetic field becomes approximately independent of ${\rm Rm}$ at ${\rm Rm}\gtrsim 10^4$, so long as ${\rm Pm}$ is not too large. In contrast, the energy-containing scale may be consistent with the field-reversal scale set by tearing, suggesting a shallow negative spectral slope in between that scale and the system scale of the turbulence. Thus, we see a degree of large-scale coherence in the amplified field that might help reconcile the view that the saturated dynamo must result in outer-scale magnetic fields~\cite{bs51,subramanian99,haugen03,haugen04,beresnyak12,XuLazarian_2016} and some previous indications, numerical and theoretical, that it produces fields whose energy resides on ${\rm Rm}$-dependent scales~\cite{cv00,mb02,scheko02_sss,maron04,Scheko04sim}. This may also be a step towards simulations becoming more consistent with Faraday-rotation observations of magnetic fields in galaxy clusters suggesting ${\gtrsim}{\rm kpc}$-scale coherence~\cite{VogtEnsslin_2005,Bonafede_2010}, as well as recent laboratory laser-plasma experiments exhibiting a ${\rm Pm}\gtrsim 1$ fluctuation dynamo~\cite{tzeferacos18,bott21}. Determining whether or not this result holds when using the pressure-anisotropic MHD~\cite{santoslima14,St-Onge2020} or kinetic~\cite{Rincon_2016,St-Onge2018} descriptions more appropriate to the weakly collisional intracluster medium awaits future work.

\begin{acknowledgments}
We thank Andrey Beresnyak and Vladimir Zhdankin for helpful discussions, as well as the two referees for comments that led to an improved presentation. This work was supported by NSF CAREER award No.~1944972, and is part of the Frontera computing project at the Texas Advanced Computing Center; it also made extensive use of the {\em Perseus} cluster at the PICSciE-OIT TIGRESS High Performance Computing Center and Visualization Laboratory at Princeton University. The work of AAS was supported in part by UK EPSRC grant EP/R034747/1.
\end{acknowledgments}

\bibliography{ref}

\end{document}